\documentclass[useAMS]{mn2e}
\usepackage{amssymb}
\usepackage{amsmath}
\usepackage{graphicx}
\usepackage[utf8]{inputenc}

\title[Hektor --- an exceptional D-type family]%
{Hektor --- an exceptional D-type family among Jovian Trojans}
\author[J. Rozehnal et al.]%
{J. Rozehnal$^{1,5}$\thanks{E-mail: rozehnal@observatory.cz (JR)}, 
M. Bro\v{z}$^{1}$, D. Nesvorn\'y$^{2}$, D. D. Durda$^{2}$, \and 
K. Walsh$^{2}$, D. C. Richardson$^{3}$, E. Asphaug$^{4}$ 
\\
\\
$^{1}$Institute of Astronomy, Charles University, Prague, V Hole\v sovi\v ck\'ach 2, 18000 Prague 8, Czech Republic\\
$^{2}$Southwest Research Institute, 1050 Walnut St., Boulder, CO 80302\\
$^{3}$Department of Astronomy, University of Maryland, College Park, 
MD 20742-2421\\
$^{4}$School of Earth and Space Exploration, Arizona State University, Tempe, 
AZ, 85287\\
$^{5}$\v Stef\'anik Observatory, Pet\v r\'in 205, 11800 Prague, Czech Republic
}

\begin{document}

\date{Accepted 2016 July 14. Received 2016 July 11; in original form 2016 
January 06}

\pagerange{\pageref{firstpage}--\pageref{lastpage}} \pubyear{2015}

\maketitle

\label{firstpage}


\begin{abstract}
In this work, we analyze Jovian Trojans in the space of suitable resonant 
elements and we identify clusters of possible collisional origin by 
two independent methods: the hierarchical clustering and a so-called 
``randombox''. Compared to our previous work (Bro\v{z} and Rozehnal 2011), we 
study a twice larger sample. 
Apart from Eurybates, Ennomos and $1996\,\rm RJ$ families, we have 
found three more clusters --- namely families around asteroids 
(20961)~Arkesilaos, (624)~Hektor in the $L_4$ 
libration zone and (247341)~$2001\,\rm UV_{209}$ in $L_5$. 
The families fulfill our stringent criteria, i.e. a high statistical 
significance, an albedo homogeneity and a steeper size-frequency distribution  
than that of background. 
In order to understand their nature, we simulate their long term collisional 
evolution with the Boulder code (Morbidelli et al. 2009) and dynamical evolution 
using a modified SWIFT integrator (Levison and Duncan, 1994). Within the 
framework of our evolutionary model, we were able to constrain the 
the age of the Hektor family to be either 1 to 4 Gyr or, less likely, 
0.1 to 2.5 Gyr, depending on initial impact geometry.
Since (624) Hektor itself seems to be a bilobed--shape body with a satellite 
(Marchis et al. 2014), i.e. an exceptional object, we address its 
association with the D--type family and we demonstrate that the moon and family 
could be created during a single impact event. We simulated the 
cratering event using a Smoothed Particle Hydrodynamics (SPH, Benz and 
Asphaug, 1994). This is also the first case of a family associated with a 
D--type parent body.

\end{abstract}




\section{Introduction}

Jovian Trojans are actually large populations of minor bodies in the 1:1 
mean motion resonance (MMR) with Jupiter, librating around $L_4$ and $L_5$ 
Lagrangian points. In general, there are two classes of theories explaining 
their origin: 
i) a theory in the framework of accretion model (e.g. Goldreich 2004, Lyra et 
al. 2009) and ii) a capture of bodies located in libration zones during a 
migration of giant planets (Morbidelli et al. 2005, Morbidelli et al. 2010, 
Nesvorn\'y et al. 2013), which is preferred in our solar system. Since the 
librating regions are very stable in the current configuration of planets and 
they are surrounded by strongly 
chaotic separatrices, bodies from other source regions (e.g. Main belt, 
Centaurs, Jupiter family comets) cannot otherwise enter the libration zones and 
Jupiter Trojans thus represent a rather primitive and isolated population. 

Several recent analyses confirmed the presence of several families among 
Trojans (e.g. Nesvorný et al. 2015, Vinogradova, 2015). The Trojan region as 
such is very favourable for dynamical studies of asteroid families, 
because there is no significant systematic Yarkovsky drift in semimajor axis due 
to the resonant dynamics. On the 
other hand, we have to be aware of boundaries of the libration zone, because 
ballistic transport can cause a partial depletion of family members.
At the same time, as we have already shown in Bro\v{z} \& Rozehnal (2011), 
no family can survive either late phases of a slow migration of Jupiter, 
or Jupiter ``jump'', that results from relevant scenarios of the Nice model
(Morbidelli et al. 2010). We thus focus on post-migration phase in this paper.

We feel the need to evaluate again our previous conclusions on even larger 
datasets, that should also allow us to reveal as-of-yet unknown 
structures in the space of proper elements or unveil possible relations between 
orbital and physical properties (e.g. albedos, colours, diameters) of Jovian 
Trojans. 

In Section \ref{sec:obsdata} we use new observational data to compute 
appropriate resonant elements. In Section \ref{sec:physchar} we use albedos 
obtained by Grav et al. (2012) to derive size-frequency distributions and  
distribution of albedos, which seem to be 
slightly dependent on the proper inclination $I_{\rm p}$. In Section 
\ref{sec:groups} we identify families among Trojans with our new ``randombox'' 
method. We discuss properties of statistically significant families in 
Section \ref{sec:props}. Then we focus mainly on the Hektor family 
because of its unique D--type taxonomical classification, which is the first of 
its kind. We also discuss its long-term dynamical evolution. In Section 
\ref{sec:coll_model} we simulate collisional evolution of Trojans and we 
estimate the number of observable families among Trojans. Finally, in Section 
\ref{sec:SPH} we simulate an origin of the Hektor family using 
smoothed-particle hydrodynamics and we compare results for single and bilobed 
targets. Section \ref{sec:conclusions} is devoted to Conclusions.
 

\section{New observational data}\label{sec:obsdata}

\begin{figure*}
\centering
\renewcommand{\tabcolsep}{0pt}
\begin{tabular}{cc}
$L_4$ Trojans & $L_5$ Trojans \\
\includegraphics[width=9cm]{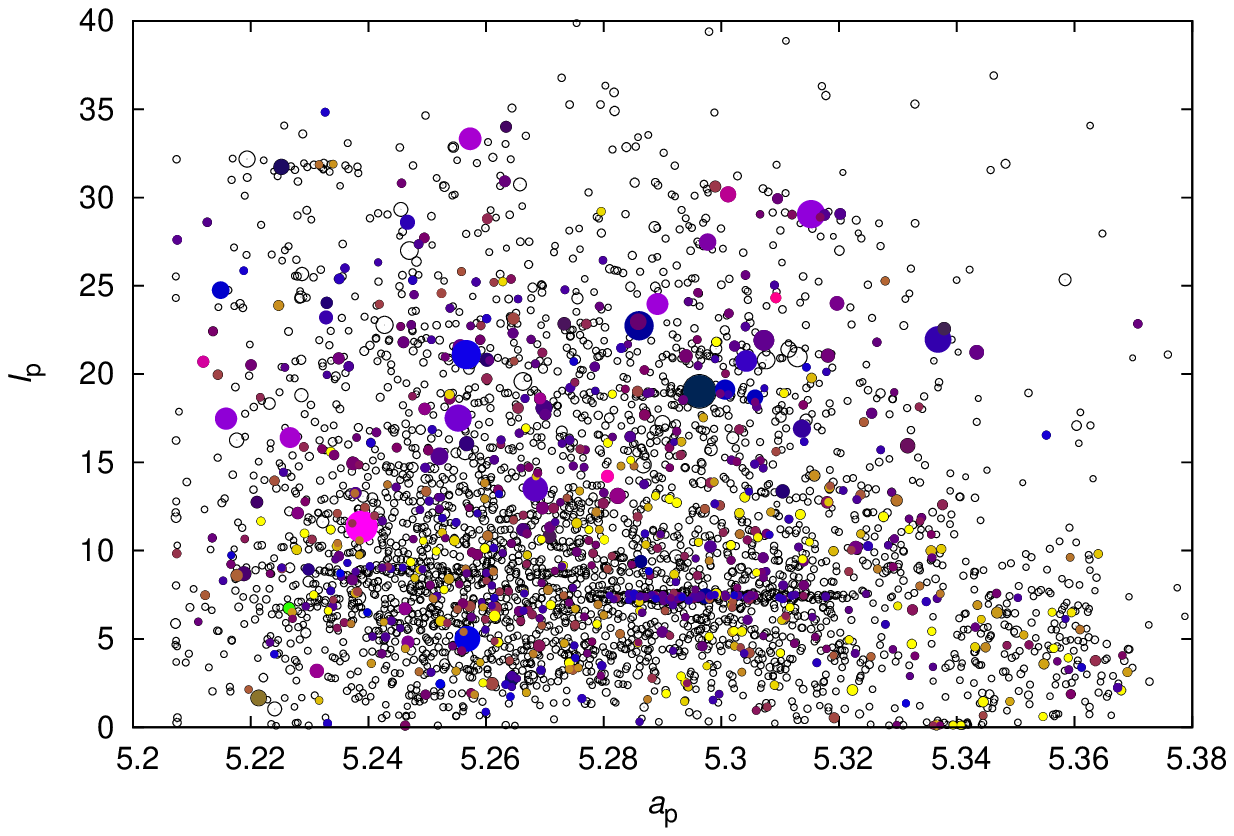} & 
\includegraphics[width=9cm]{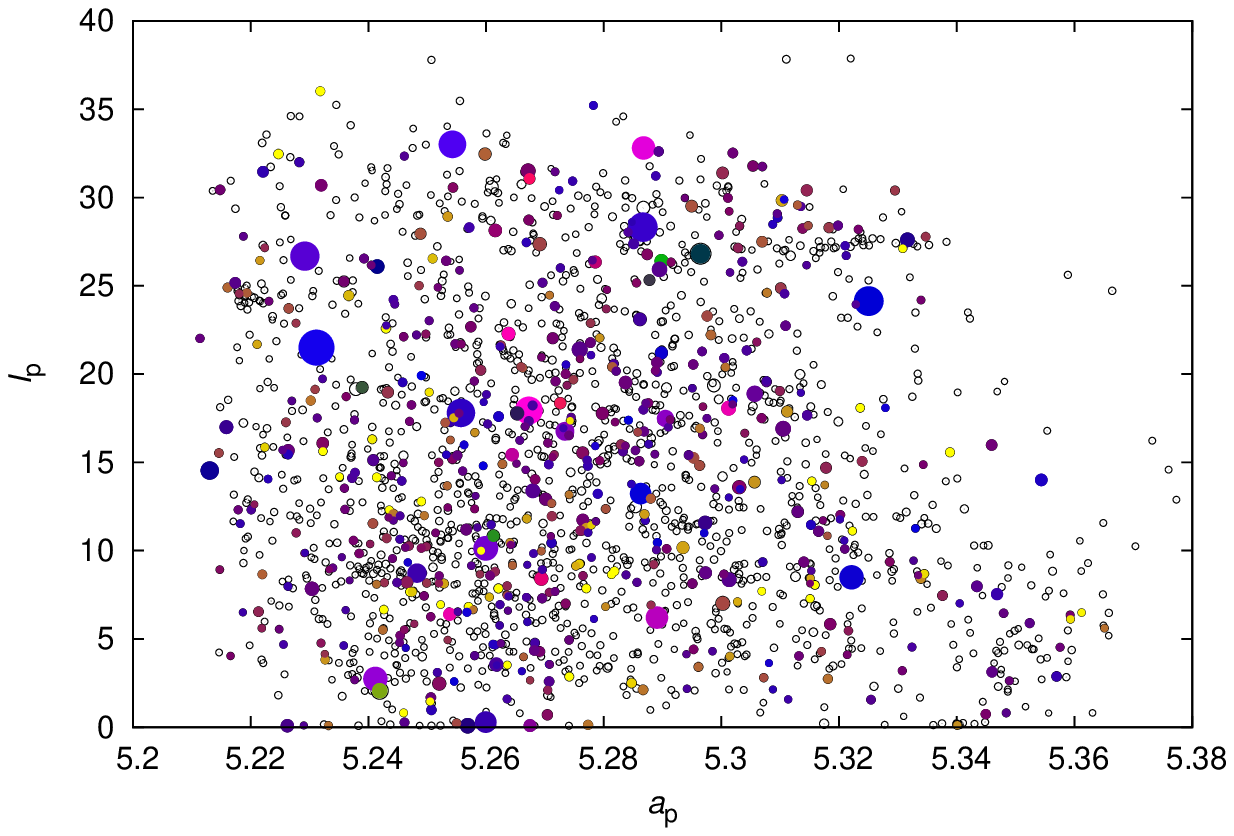} \\
\includegraphics[width=9cm]{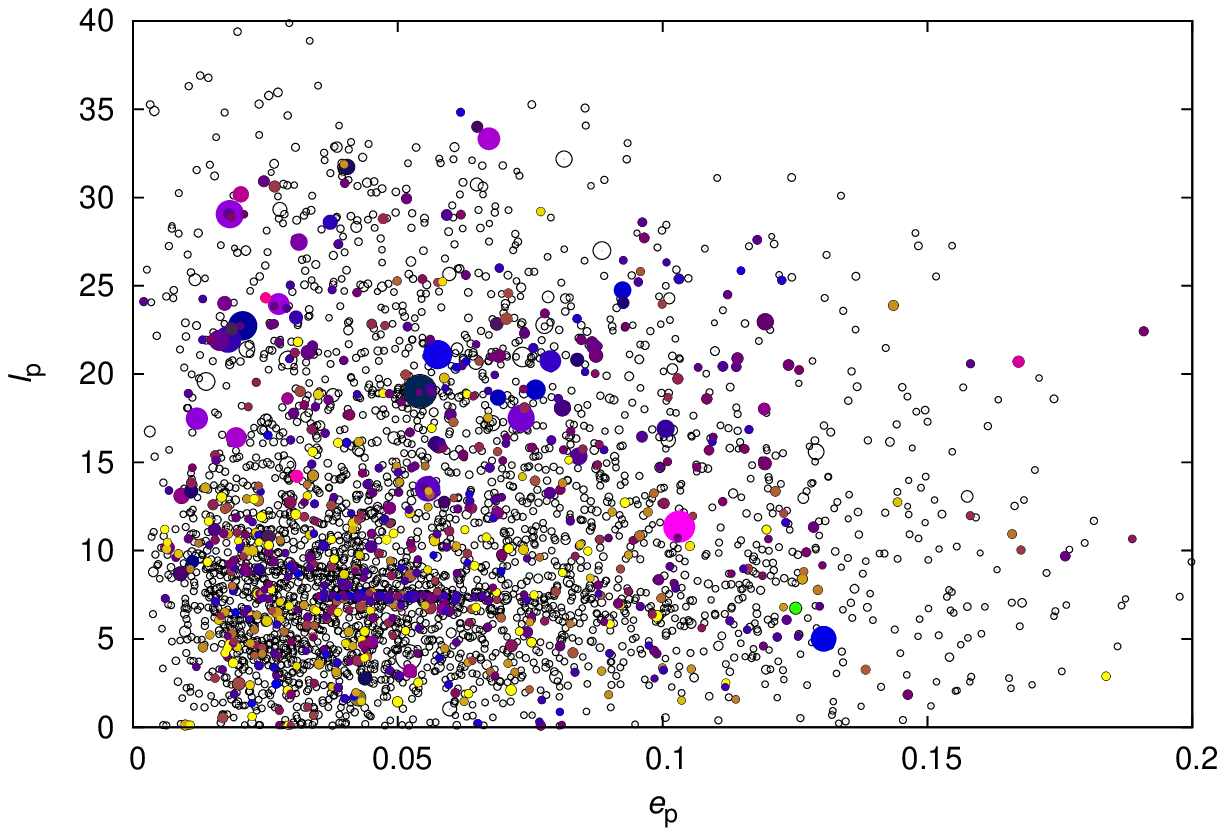} & 
\includegraphics[width=9cm]{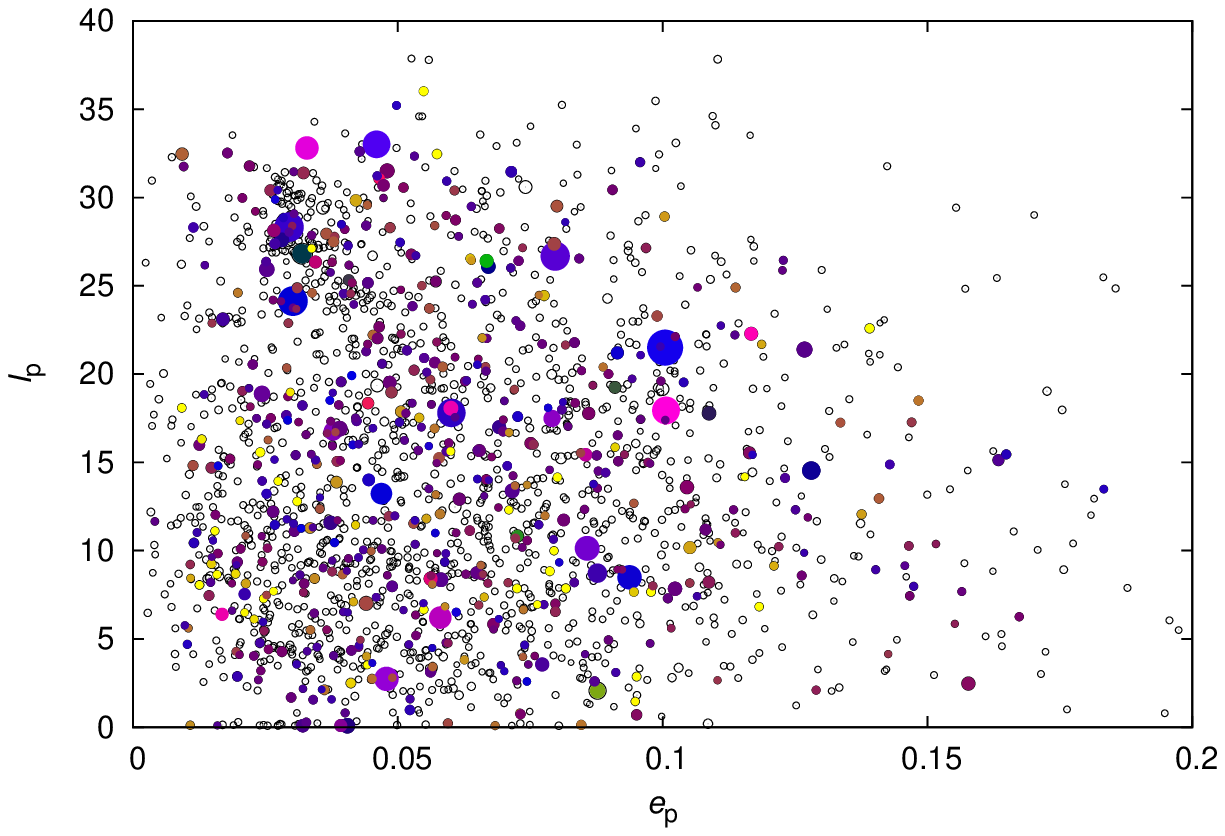} \\
\end{tabular}
\caption{
The resonant semi-major axis 
vs inclination $(a_{\rm{p}}, I_{\rm{p}})$ 
(top) and eccentricity vs inclination $(e_{\rm{p}}, I_{\rm{p}})$ (bottom) for 
$L_4$ (left) and $L_5$ Trojans (right).
The circles indicate relative diameters of bodies, as determined by WISE (Grav 
et al. 2011), or when unavailable, computed from the absolute magnitude~$H$ 
and geometric albedo~$p_{\rm V}$, which we assumed to be $p_{\rm V} = 0.07$ for 
both the $L_4$ and $L_5$ Trojans (WISE median value is $p_{\rm V} = 0.072$ for 
$L_4$ and  $p_{\rm V} = 0.069$ for $L_5$ Trojans). Colours correspond to the 
values of $p_{\rm V}$, blue are dark ($p_{\rm V} \simeq 0.05$) and yellow are 
bright ($p_{\rm V} \simeq 0.25$). One can see clearly all asteroid families on 
this plot, especially in $(a_{\rm{p}}, I_{\rm{p}})$, because they tend to be 
confined in inclinations.}
\label{trojans_L4_arer_sizes}
\end{figure*}

\subsection{Resonant elements}

We computed resonant elements, i.e. the averaged semimajor axis 
$\bar{a}$, libration amplitude $\Delta a_{\rm p}$, 
eccentricity $e_{\rm p}$ and inclination $I_{\rm p}$ of 3907 Trojans in $L_4$ 
cloud and 1945 Trojans in $L_5$ cloud. As an input, we used osculating elements 
listed in AstOrb catalogue (Bowell et al. 2002), released in July 2014. A 
detailed description of the resonant elements computation can be found in 
Bro\v{z} \& Rozehnal (2011). 
Positions of Trojans in the space of proper elements ($a_{\rm p}$, $I_{\rm 
p}$), where $a_{\rm p} = \bar{a} + \Delta a_{\rm p}$, and ($e_{\rm 
p}$, $I_{\rm p}$), calculated with a suitably modified 
version of the SWIFT integrator (Levison \& Duncan, 1994), are presented 
graphically in Figure \ref{trojans_L4_arer_sizes}, together with their sizes 
and albedos. 
\footnote{The table of resonant elements is listed online at 
\tt{http://sirrah.troja.mff.cuni.cz/$\sim$mira/mp/trojans/}.}

\subsection{WISE and AKARI albedos and diameters}
To construct size-frequency distributions of the whole $L_4$ and $L_5$ Trojan 
populations and later of individual families, we mostly used WISE albedos and 
diameters derived by Grav et al. (2012). We also compared the respective values 
to AKARI, as reported by Usui et al. (2011).\footnote{While there are some 
differences between individual values even at $3\sigma$ level, they do not seem 
to be important for population studies like ours.} 

We used albedo values of 1609 Trojans in both $L_4$ and $L_5$ clouds 
obtained by WISE; about one third of these albedos were obtained during 
cryo-phase, the rest were measured in post-cryo-phase (see Grav et al., 2011).


\section{Physical characterisation of Trojan populations}\label{sec:physchar}

\subsection{Albedo distribution and taxonomy}

The values of visible albedos $p_{\rm V}$ of Trojans derived by Grav et 
al. (2012) vary in the range from $p_{\rm V} = 0.025$ to $p_{\rm V} \simeq 
0.2$. Distributions of 
albedos are qualitatively the same for both $L_4$ and $L_5$ populations. 
The median albedo of WISE sample is $\widetilde{p}_{\rm v}=0.072 \pm 0.017$ for 
$L_4$ and $\widetilde{p}_{\rm v}=0.069 \pm 0.015$ for $L_5$. These values of 
visible albedos mostly correspond to C or D taxonomical classes in Tholen 
taxonomic classification scheme (Mainzer et al., 2011). However, there is a 
significant presence of small asteroids $(D<15\,\rm km)$ with apparently high 
albedo --- almost 20\,\% of asteroids in $L_4$ and 13\,\% of asteroids in $L_5$ 
have albedo $p_{\rm V} > 0.10$. As stated in Grav et al. (2012), this is 
probably not a physical phenomenon, it is rather due to the fact that for 
small diameters the photon noise contribution becomes too significant. 

When we compute the median albedo from AKARI data, we realize that its value is 
slightly lower ($\widetilde{p}_{\rm v}=0.054 \pm 0.005$) than that from WISE, 
but when we compute the median from WISE values for the same asteroids which 
are listed in AKARI catalogue, we obtain a similar value ($\widetilde{p}_{\rm 
v}=0.061 \pm 0.012$). What is more serious, AKARI and WISE data differ 
considerably for large asteroids with $D>100\,\rm km$ --- the average 
difference between albedos is $|p_{\rm V_{AKARI}} - p_{\rm V_{WISE}}| = 0.02$. 
The same difference we see in derived diameters. These discrepancies may be 
caused for example by limitations of the thermal model (cf. spheres in NEATM 
models). Hereinafter, we prefer to use the WISE data when available, because 
they represent orders of magnitude larger sample than AKARI. 

\begin{figure}
\centering
\includegraphics[width=8.5cm]{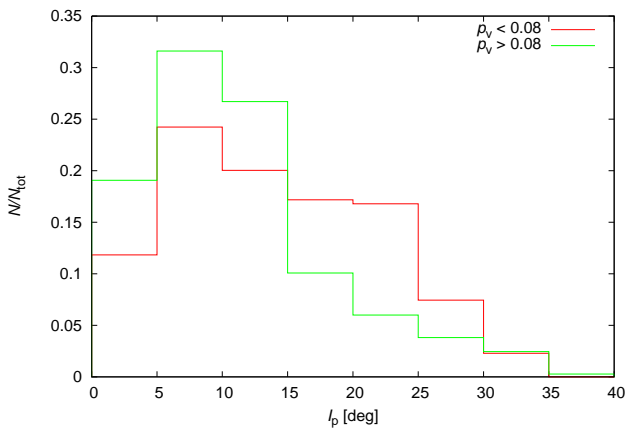}
\caption{The differential histogram of the resonant inclination $I_{\rm p}$ for 
$L_4$ Trojans with a lower albedo ($p_{\rm V} < 0.08$, red) and a higher one 
($p_{\rm V} >0.08$, green). Eurybates family was removed from the dataset.}
\label{albedo_separation_L4_hist}
\end{figure}

When we split Trojan asteroids according to their albedo into two 
rather artificial subpopulations with $p_{\rm V} < 0.08$ and $p_{\rm V} > 0.08$ 
respectively, and then we compute distributions of these 
subpopulations with respect to the resonant inclination $I_{\rm p}$, we get two 
different pictures. As can be seen in Figure \ref{albedo_separation_L4_hist}, 
most bodies have resonant inclinations $I_{\rm p} < 15^{\circ}$, but there 
are 77\,\% of bodies with higher albedo with $I_{\rm p} < 15^{\circ}$, while 
only 55\,\% of the population with lower albedo is located in the same range of 
inclinations. This is a similar phenomenon as described by Vinogradova (2015), 
who reported different upper limits in inclinations for different taxonomical 
types obtained mostly from SDSS colour data.

\subsection{Size-frequency distributions}\label{sec:SFDs}
\begin{figure*}
\centering
\renewcommand{\tabcolsep}{0pt}
\begin{tabular}{cc}
$L_4$ Trojans & $L_5$ Trojans \\
\includegraphics[width=8.5cm]{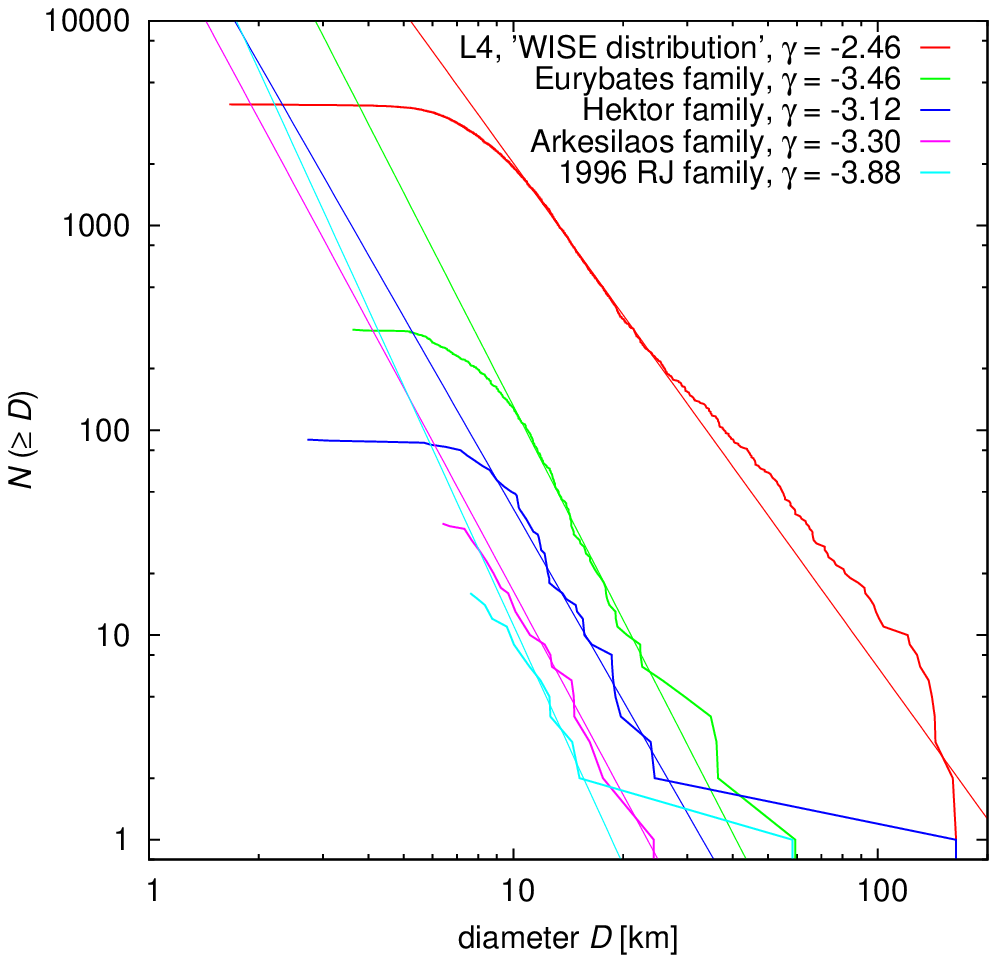} & 
\includegraphics[width=8.5cm]{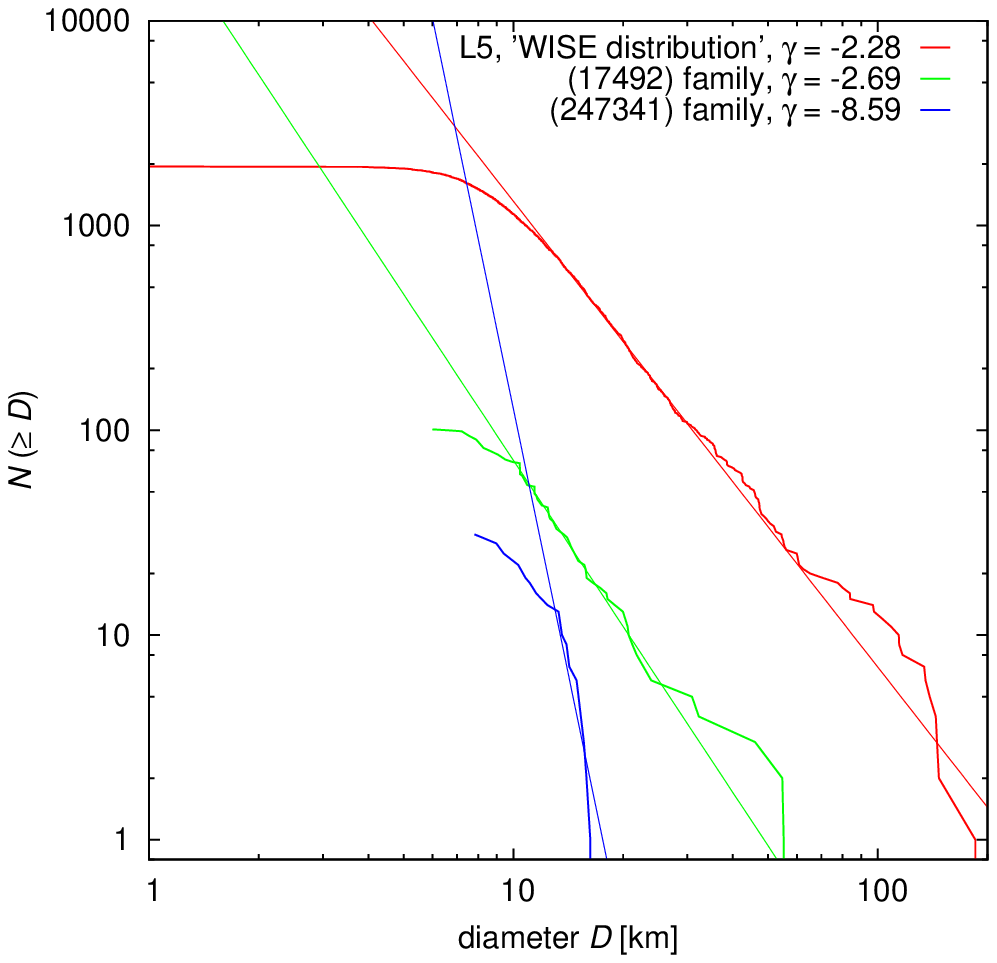} \\
\end{tabular}
\caption{Size-frequency distributions for both $L4$ and $L_5$ Trojans, 
constructed using the albedos measured by WISE satellite (Grav et al. 2012). 
Since WISE data cover just about 18\,\% of $L_4$ and 29\,\% of $L_5$ Trojans 
known today, we assigned albedos randomly from the WISE distribution to the 
remaining Trojans. We also present SFDs of individual asteroid families 
discussed 
in the main text. There are also our fits of each SFD in the range $D=12$ to 30 
km by the power law $N(>D) = C D^\gamma$. As we can see, both clouds seem to be 
near the collisional equilibrium ($\gamma\simeq-2.5$, Dohnanyi 1969), while 
most families have slope $\gamma$ significantly steeper. Of course, we can 
expect the slopes of the SFDs become shallower for smaller $D$ due to 
observational incompleteness.}
\label{SFD_WISE}
\end{figure*}

The WISE data (Grav et al. 2011, 2012) provide very useful source of 
information 
on diameters we need to construct size-frequency distributions (SFDs) of Trojan 
populations in $L_4$ and $L_5$. However, the sample measured by WISE is not 
complete. In our previous work (Bro\v{z} and Rozehnal, 2011), we constructed 
the SFDs assuming a constant albedo which we set to be equal to the 
median albedo of Trojans that was measured back then. Since the number of 
measurements was very low (several tens), this was the only reasonable way. 
Now we choose another method to construct more reliable SFDs. As we calculated 
resonant elements for more than 5800 Trojans and we have more than one quarter 
of appropriate albedos, we constructed the SFDs by assigning albedos randomly 
from the observed WISE distribution to the remaining Trojans, whose albedo was 
not measured. To avoid a bias, we compared different SFDs constructed with 
different random generator seeds and we realized that the overall shape of 
SFDs does not change noticeably, the slope $\gamma$ varies in the range of 
$\pm 0.1$ at most. The SFDs we constructed this way are shown in Figure 
\ref{SFD_WISE}.

The SFDs for the $L_4$ and $L_5$ clouds look slightly different, 
especially in the size range from 60\,km to 100\,km. This part of the SFD is not 
influenced by the Eurybates family, the largest family among Trojans, because 
all 
its members have diameters $D<50\,\rm{km}$. We used these SFDs to determine the 
ratio of the number of asteroids in $L_4$ and $L_5$ clouds. There are 2746 
asteroids with diameter $D>8\,\rm km$ in $L_4$ and 1518 asteroids in $L_5$. When 
we remove all family members with diameters $D>8\,\rm km$, we have 2436 
asteroids in~$L_4$ and 1399 in~$L_5$. However, this sample may be still 
influenced by debris produced by catastrophic disruptions of small bodies 
($D\geq 50\,\rm km$), which need not to be seen as families. Counting only 
asteroids with diameter $D>20\,\rm km$, which 
corresponds to the absolute magnitude $H\simeq12$, and removing family members, 
we get the ratio $N_{L_4}/N_{L_5}=1.3 \pm 0.1$. As this is entirely consistent 
with value of Nesvorn\'{y} et al. (2013), which was derived for Trojans with 
$H>12$, and with Grav et al. (2012), whose estimate is $N_{L_4}/N_{L_5} = 1.4 
\pm 0.2$, we can confirm a persisting asymmetry between the number of $L_4$ and 
$L_5$ Trojans in new data. Although for bodies with diameter 
$D>100\,\rm km$, the $L_5$ cloud has 
more asteroids than $L_4$, the total number of these bodies is of the 
order of 10, so this is just an effect of small-number statistics and does 
not affect the $N_{L_4}/N_{L_5}$ ratio much.

\begin{table*}\vspace{-5mm}\centering
\caption{Physical properties of Trojan families identified by both 
the ``randombox'' and the HCM methods. We list Family Identification Number 
(FIN, as in Nesvorn\'{y} et al., 2015), the designation of the family, 
the designation of the asteroid with which the family is associated (i.e. 
usually the largest remnant of the parent body), the cutoff velocity $v_{\rm 
cutoff}$, for which family is still clearly detached from the background, and 
the number of members of the family $N_{\rm{memb}}$ corresponding to 
the respective $v_{\rm cutoff}$. Next we list optical albedos $p_{\rm V}$ of 
associated bodies determined by Grav et al. (2012) from WISE observations, and 
their taxonomical classification.} 
\small
\begin{tabular}{c|c|c|c|c|c|c|c|}	
\hline\
FIN & family designation & cloud & asteroid designation & $v_{\rm cutoff} \rm 
[m\,s^{-1}]$ & $N_{\rm{memb}}$ & $p_{\rm V} \rm (WISE)$ & tax. 
type \\
\hline\hline
004 & Hektor & $L_4$ & (624) Hektor & 110 & 90 & $0.087 \pm 0.016$ & D \\
005 & Eurybates & $L_4$ & (3548) Eurybates & 60 & 310 & $0.060 \pm 0.016$ & C/P 
\\
006 & $1996\,\rm RJ$ & $L_4$ & (9799) $1996\,\rm RJ$ & 140 & 17 & $0.082 \pm 
0.014$ & 
-- \\
008 & Arkesilaos & $L_4$ & (20961) Arkesilaos & 55 & 35 & n/a & -- \\
009 & Ennomos & $L_5$ & (17492) Hippasos & 100 & 104 & $0.064 \pm 0.012$ & -- \\
010 & $2001\,\rm UV_{209}$ & $L_5$ & (247341) $2001\,\rm UV_{209}$ & 120 & 30 & 
$0.088 
\pm 0.023$ & -- \\
\hline
\vspace{2mm} 

\end{tabular}
\label{fams_list}
\end{table*}

\begin{table*}\vspace{-5mm}\centering
\caption{Derived properties of Trojan families. We list here the family 
designation, the diameter of the largest remnant $D_{\rm LR}$, the minimal 
diameter of the parent body min $D_{\rm PB}$, 
obtained as the sum of all observed family members, the diameter of the parent 
body $D_{\rm PB(SPH)}$ and the mass ratio $M_{\rm{LR}}/M_{\rm{PB}}$ of 
the largest fragment and the parent body, both derived from our fits by scaled 
SPH simulations performed by Durda et al. (2007). We use this ratio to 
distinguish between the catastrophic disruption ($M_{\rm{LR}}/M_{\rm{PB}} < 
0.5$) and the cratering ($M_{\rm{LR}}/M_{\rm{PB}}>0.5$).
Finally, there is the escape velocity $v_{\rm esc}$ from the parent 
body and estimated age of the 
family derived in this and our previous work (Bro\v{z} and Rozehnal, 
2011).} 
\small
\begin{tabular}{|c|c|c|c|c|c|c|c|}	
\hline\
family desig. & $D_{\rm LR}$ [km] & min $D_{\rm PB}$ & $D_{\rm PB(SPH)}$ & 
$M_{\rm{LR}}/M_{\rm{PB}}$ & $v_{\rm esc} \rm [m\,s^{-1}]$ & age [Gyr] & notes, 
references\\
\hline\hline
Hektor & $250 \pm 26$ & 250 & 257 & 0.92 & 73 & 0.3 or 3 & 
1, 3\\
Eurybates & $59.4 \pm 1.5$ & 100 & 155 & 0.06 & 46 & 1.0 to 3.8 & 2\\
$1996\,\rm RJ$ & $58.3 \pm 0.9$ & 61 & 88 & 0.29 & 26 & -- & 
2,4\\
Arkesilaos & $24 \pm 5$ & 37 & 87 & 0.02 & 16 & -- & 2\\
Ennomos & $55.2 \pm 0.9$ & 67 to 154 & 95 to 168 & 0.04 to 0.19 & 29 to 66 & 1 
to 2 & 2, 5\\
$2001\,\rm UV_{209}$ & $16.3 \pm 1.1$ & 32 & 80 & 0.01 & 14 & -- & 2\\
\hline
\end{tabular}

\scriptsize{$^1 D_{\rm LR}$~derived~by~Marchis~et~al.~(2014)}, 
\scriptsize{$^2 D_{\rm LR}$ derived by Grav et al. (2012)},
\scriptsize{$^3$ bilobe, satellite (Marchis et al. 2014)},
\scriptsize{$^4$ very compact, Bro\v{z} and Rozehnal (2011)},
\scriptsize{$^5 D_{\rm PB}$ strongly influenced by interlopers},
\scriptsize{$^6$ The largest fragment of Ennomos family is (17492) Hippasos}.
\label{fams_prop}
\end{table*}

\section{Families detection methods}\label{sec:groups}
A brief inspection of the resonant-element space ($a_{\rm p}$, $e_{\rm 
p}$, $I_{\rm p}$) (see Figure \ref{trojans_L4_arer_sizes}), 
reveals several locations with higher concentrations of bodies. These could be 
collisional families, created by a disruption of a parent body during 
a random collision, but they could also originate randomly by chaotic diffusion 
and due to effects of secular and high-order resonances. To be regarded as a 
family, the cluster must comply with, inter alia, the following criteria: i) 
it must be concentrated in the space of proper elements; ii) the cluster must 
have the SFD different from that of the whole $L_4$ and $L_5$ population; iii) 
the last criterion is usually spectral, or at least, albedo homogeneity of 
family members, but so far, there are not enough sufficiently accurate data for 
Trojans, especially for bodies with diameters $D< 50\,\rm km$, which usually 
form a substantial part of Trojan families. Therefore we cannot perform any 
detailed spectral analysis in this work. 

We analyzed the space of resonant elements both in terms of mutual distances 
among bodies and in terms of statistical probability that clusters 
are \textit{not} random.

\subsection{Randombox method}

Besides the commonly used hierarchical clustering method (HCM, Zappal\`{a} et 
al., 1990), we applied a ``randombox'' method, based on numerical Monte-Carlo 
simulations. This method allows us to compute the statistical significance of 
the clusters, i.e. the probability that the cluster is a random concentration 
of bodies in the space of proper elements ($a_{\rm p}, e_{\rm p}, \sin{I_{\rm 
p}}$).

\begin{figure*}
\renewcommand{\tabcolsep}{0pt}
\renewcommand{\arraystretch}{0.2}
\centering
\begin{tabular}{cc}
$L_4$ Trojans & $L_5$ Trojans \\
\\
\includegraphics[width=8.9cm]{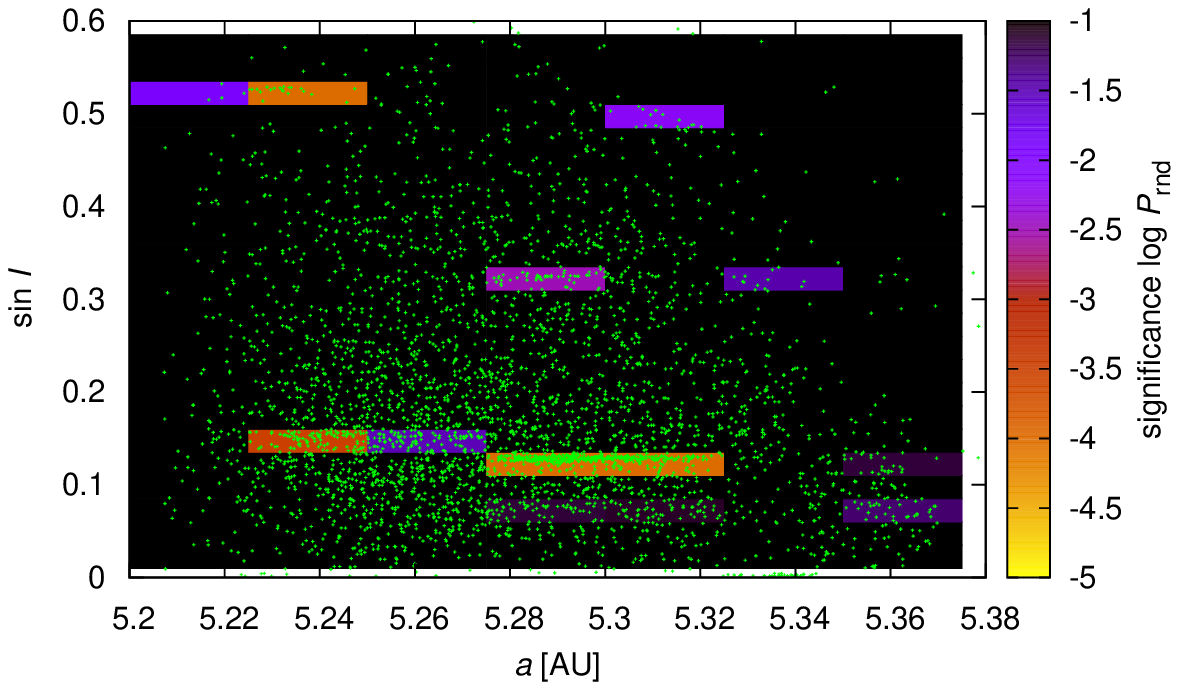} & 
\includegraphics[width=8.9cm]{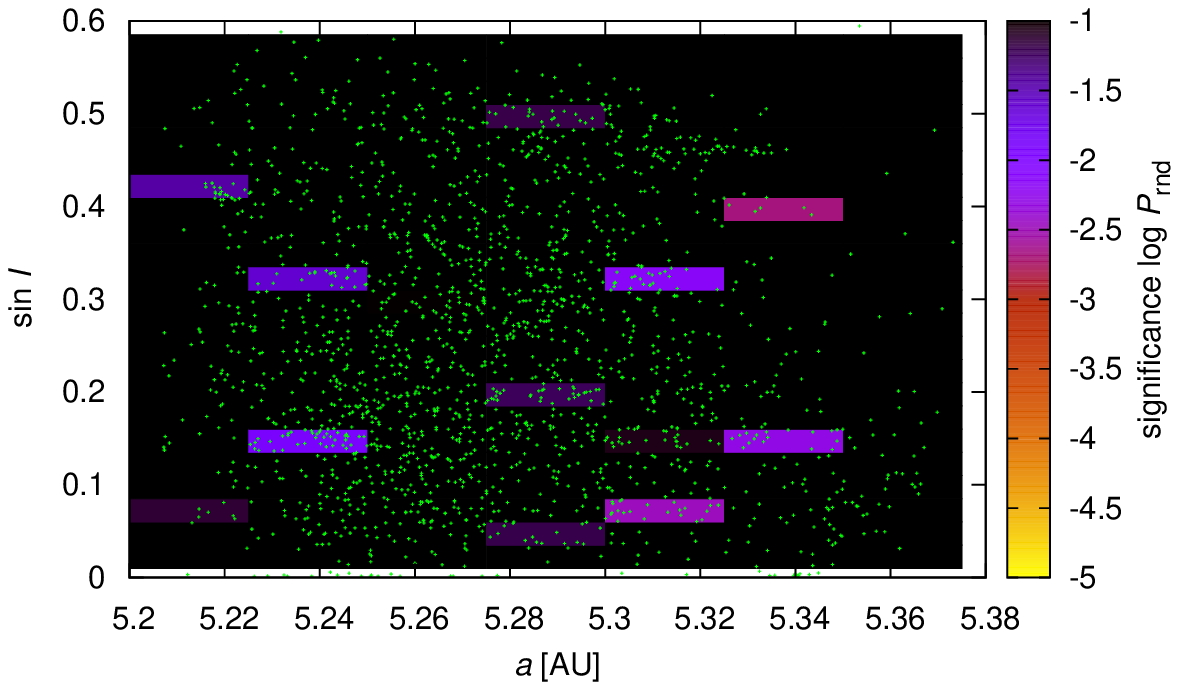} \\
\end{tabular}
\caption{The statistical significance $p$ expressed as colour on the 
logarithmic scale for observed asteroids in the proper semimajor axis vs proper 
inclination plane ($a_{\rm{p}},\sin I_{\rm{p}}$) (i.e. the same data as in 
Figure \ref{trojans_L4_arer_sizes}). $L_4$ Trojans are on the left, $L_5$ 
Trojans on the right. We computed the values of $p$ for 7 times 18 boxes using 
our ``randombox'' method 
The range in proper eccentricity is 0.00 to 0.20. Statistically significant 
groups appear as orange boxes and they correspond to the families reported in 
Table \ref{fams_list}.}
\label{Randombox}
\end{figure*}

We divided the space of proper elements into equally sized ``boxes'' with 
dimensions $\Delta a_{\rm p} = 0.025\, \rm au$, $\Delta e_{\rm p} = 0.2$ and 
$\Delta \sin{I_{\rm p}} = 0.025$. Then we created $N = 100,000$ random 
distributions of the same number of bodies which are observed together in 
the given box and two adjacent boxes (in the direction of the 
$y$-axis, cf. Figure \ref{Randombox}), and we counted number of positive 
trials $N^+$, for which the randomly generated number of bodies in the central 
box was larger than the observed one. From here we can calculate the 
probability $P_{\rm rnd}$, that the observed number of bodies in the box is 
random: $P_{\rm rnd} = N^+ / N$.  

Alternatively, one can also use our analytical formula:
\begin{equation}\label{randombox_p}
p_{\rm rnd} = \frac{\sum _{k=n_2} ^n C(n,k)V^\prime (n_{\rm 
box}-1,n-k)}{V^\prime(n_{\rm box},n)}\,,
\end{equation}
where $n$ denotes the total number of bodies, $n_{\rm box}$ is the total number 
of boxes (3 in our case), $n_2$ is the observed number of bodies in 
the middle box, $k$ is the number of observed bodies in the current box, 
$C(n,k)$ are combinations without repetitions, i.e. the total number of trials 
to select $k$ bodies observed in the current box from the total number of $n$ 
bodies; $V^\prime (n_{\rm box}-1,n-k)$ are variations with repetitions, 
i.e. the total number of trials to distribute the remaining bodies into the 
remaining boxes; and $V^\prime(n_{\rm box},n)$ are also variations with 
repetitions, i.e. the total number of trials to distribute all $n$ bodies into 
all $n_{\rm box}$ boxes. We verified the results of the analytical formula 
(\ref{randombox_p}) by the MC method.

We plot the results in Figure \ref{Randombox} for both the $L_4$ and $L_5$ 
clouds. In comparison with Figure \ref{trojans_L4_arer_sizes}, one can see that 
for all clusters we identified as families the probability $P_{\rm rnd}$ varies 
between $2\cdot 10^{-3}$ and $5\cdot 10^{-5}$, i.e. the probability that 
clusters are random fluctuations is indeed very low.

We also re-evaluated all families identified by the hierarchical clustering 
method using the ``randombox'' method, which makes our decision whether the 
cluster is a real family much more quantitative.  

\subsection{Hierarchical clustering method}\label{sec:HCM}
We also used the HCM independently to extract significant clusters. Families 
identified by both the ``randombox'' and HCM methods are listed in 
Table \ref{fams_list}. For each family, we constructed a dependence of 
the number of members of the cluster $N_{\rm{memb}}$ on the cutoff velocity 
$v_{\rm{cutoff}}$. Because the number of members of a real collisional family 
rises first slowly with rising $v_{\rm{cutoff}}$ (Bro\v{z} and Rozehnal, 2011) 
--- in contrast with random clusters which are merging very 
quickly with the background --- the constructed dependence allows us to 
guess a realistic number of family members $N_{\rm{memb}}$. For all families 
listed in Table \ref{fams_list} we were convinced that they fulfill this 
criterion. However, we cannot distinguish possible interlopers this way, and 
it is also possible that some fraction of family members with high 
$v_{\rm{cutoff}}$ (so called halo, as in Bro\v{z} and Morbidelli, 2013) remains 
unidentified in the surrounding background.

\section{Properties of statistically significant families}\label{sec:props}

\subsection{Eurybates}

As we have already demonstrated in Bro\v{z} and Rozehnal (2011), the family 
associated with asteroid (3548) Eurybates is the largest collisional family, 
and it is the only family among Trojans with the parent body size 
$D_{\rm{PB}} > 100\,\rm{km}$, which originated by a catastrophic disruption 
(this means that the mass ratio of the largest remnant to the parent body 
$M_{\rm{LR}}/M_{\rm{PB}}< 0.5$). 

Using new albedos derived by Grav et al. (2012), we recalculated the overall 
SFD slope of the family to be $\gamma = -3.4 \pm 0.1$. As the WISE sample 
provides albedos for only about 1/5 of the family members, we calculated two 
values of~$\gamma$: the first one assuming that remaining asteroids have 
a constant albedo $p_{\rm V} = 0.06$, the second one by assigning albedos 
randomly from the WISE distribution, as described in Section \ref{sec:SFDs}. 
Both values are equal within their errorbars. The new slope $\gamma$ is 
significantly steeper than our previous calculation ($\gamma = -2.5 \pm 0.1$), 
derived with the assumption of a constant albedo of all members of the family. 
The lower value was most likely caused by a significant observational 
incompleteness in the size  range from $D=12\,\rm km$ to $D=30\,\rm km$.

We also derived the new value of the parent body diameter, which is still above 
the limit of 100 km. An extrapolation of the SFD by a power law gives the value 
$D_{\rm{PB}}\simeq 140\,\rm{km}$. By fitting the synthetic SFDs from SPH 
simulations (Durda et al., 2007), we obtained the value 
$D_{\rm{PB(SPH)}}\simeq155\,\rm{km}$.      

\subsection{Hektor --- the first D--type family}\label{subsec:Hektor}

Since asteroid (624) Hektor is a close binary with a satellite (Marchis et al. 
2014), i.e. an exceptional object, we want to address its association with the 
family. The cluster around the largest Trojan asteroid appears in the space of 
proper elements as a relatively compact group, which is limited particularly in 
proper inclinations, $I_{\rm{p}}\in\langle 18.13^\circ;19.77^\circ\rangle$, and 
with resonant semimajor axes located in the interval $a_{\rm{p}}\in\langle 
5.234;5.336\rangle$ au. The number of members of this group 
slowly increases with increasing cutoff velocity up to
$v_{\rm{cutoff}} \simeq 110\,\rm{m}\,\rm{s^{-1}}$, above which it quickly 
joins the background. With our randombox method, we estimated the probability 
that the family is just a random fluke to be as low as  
$P_{\rm{rnd}}\simeq2\cdot10^{-3}$. 

The nominal diameter of asteroid (624) Hektor derived from its albedo is 164 km 
(Grav et al., 2012), but the albedo measured by AKARI $p_V = 0.034 \pm 0.001$ 
(Usui et al., 2011) totally differs from that measured by WISE, $p_V = 0.087 
\pm 0.016$. and these values do not match even within the error limits. This 
may be caused by applying a thermal model assuming spheres to the bilobed shape 
of the asteroid (Marchis et al., 2014). We hence do not determine Hektor's 
diameter from its albedo, but from fits of Marchis et al. (2014), which 
effective value $D = (250 \pm 26)\,\rm km$ is suitable within its uncertainty 
for all possible geometries (convex, bilobe and binary). For other bodies in 
family we use a nominal value $p_{\rm V}=0.072$, which is the median of WISE 
measurements.

Asteroid (624) Hektor is often classified as D-type
(e.g. Cruikshank et al., 2001, Emery et al., 2006,
Emery et al., 2011). We tried to evaluate taxonomical
classification of other family members and we have
found colours for two more expected family members
in SDSS-MOC vers. 4 (Ivezic et al., 2002):
asteroids (65000) 2002 AV63 and (163702) 2003 FR72.
Even though the photometric noise in individual bands is not negligible
($\sigma_i = 0.02\,{\rm mag}$ up to $\sigma_u = 0.12\,{\rm mag}$)
both of them are D-types, with principal components
(aka slopes) ${\rm PC}_1 > 0.3$. This seems to support
the D-type classification of the whole family.

We also tried to constrain the taxonomic classification
of the family members by comparing their infrared albedos $p_{\rm IR}$
and visual albedos $p_{\rm V}$ as described in Mainzer et al. (2011),
but there are no data for family members in the W1 or W2 band of the WISE 
sample, which are dominated by reflected radiation. 

The fact that we observe a collisional family associated with a 
D--type asteroid is the main reason we use word ``exceptional'' in connection 
with the Hektor family. As we claimed in Bro\v{z} et al. (2013), in all regions 
containing a mixture of C--type and D--type asteroids (e.g. Trojans, Hildas, 
Cybeles), there have been only C--type families observed so far, which could 
indicate that disruptions of D--type asteroids leave no family behind, as 
suggested by Levison et al. (2009). Nevertheless, our classification of 
the Hektor family as D--type is not in direct contradiction with this 
conclusion, because Levison et al. (2009) were concerned with catastrophic 
disruptions, while we conclude below that the Hektor family originated from a 
cratering event, i.e. by an impactor with kinetic energy too small to disrupt 
the parent body.

\subsubsection{Simulations of long-term dynamical evolution}\label{sub:dyn_evol}

To get an upper limit of the age of the Hektor family, we simulated a 
long-term evolution of \textit{seven synthetic families} created for different 
breakup geometries. Our model included four giant planets on current orbits, 
integrated by the symplectic integrator  SWIFT (Levison and Duncan, 1994), 
modified according to Laskar and Robutel (2001), with the time step of $\Delta t 
= 91\,\rm days$ and time span 4 Gyr. 

We also accounted for the Yarkovsky effect in our simulations. 
Although in a first-order theory, it is not effective in zero-order 
resonances (it could just shift libration centre, but there is no systematic 
drift in semimajor axis) and the observed evolution of proper elements is mainly 
due to chaotic diffusion, in higher-order theories the Yarkovsky effect can 
play some role. In our model, we assumed a random distribution of spins and 
rotation periods (typically several hours), the bulk and surface density 
$\rho_{\rm{bulk}} = \rho_{\rm{surf}} = 1.3\,\,\rm{g}\,\rm{cm^{-3}}$, the 
thermal conductivity $K = 0.01\,\,\rm{W}\,\rm{m^{-1}}\,\rm{K^{-1}}$, 
the specific heat capacity $C = 680\,\,\rm{J}\,\rm{kg^{-1}}\,\rm{K^{-1}}$, 
the Bond albedo $A_{\rm B} = 0.02$ and the IR emissivity $\epsilon = 0.95$.

We created each synthetic family by assigning random velocities to 234 
bodies (i.e. 3 times more than the number of the observed family members), 
assuming an isotropic velocity field with a typical velocity of 
$70\,\,\rm{m}\,\rm{s^{-1}}$, corresponding to the escape velocity from parent 
body (Farinella et al., 1993). Here we assumed the velocity of 
fragments to be size independent. Possible trends in the ejection velocity 
field cannot be easily revealed in the ($a, H$) space in the case of the Hektor 
family, because of its origin by a cratering event -- there is a large gap in 
the range between absolute magnitude of (624) Hektor ($H = 7.20$) and other 
bodies ($H > 11.9$), so we are not able to distinguish a simple Gaussian 
dispersion from the physical dependence (cf. Carruba et al. 2016). Either way, 
we are interested in the orbital distribution of \textit{mostly} small bodies. 
Our assumption of size-independent ejection velocity is also in good agreement 
with results of SPH models (see Subsection \ref{subsec:velocity_fields} and 
Figure \ref{fig:SPH_v_evol}).

To create a synthetic family in the same position as occupied by the observed 
Hektor family, we integrated the orbit of asteroid (624) Hektor with osculating 
elements taken from AstOrb catalogue (Bowell et al., 2002), until we got 
appropriate values of the true anomaly $f$ and the argument of pericentre 
$\omega$. We tried values of $f$ ranging from $0^\circ$ to $180^\circ$ with the
step of $30^\circ$ and $\omega$ always satisfying the condition $f + \omega = 
60^\circ$, i.e. we fixed the angular distance from the node to ensure a 
comparably large perturbations in inclinations. 

\begin{figure*}
\centering
\renewcommand{\tabcolsep}{0pt}
\begin{tabular}{ccc}
\\
\includegraphics[width=6cm]{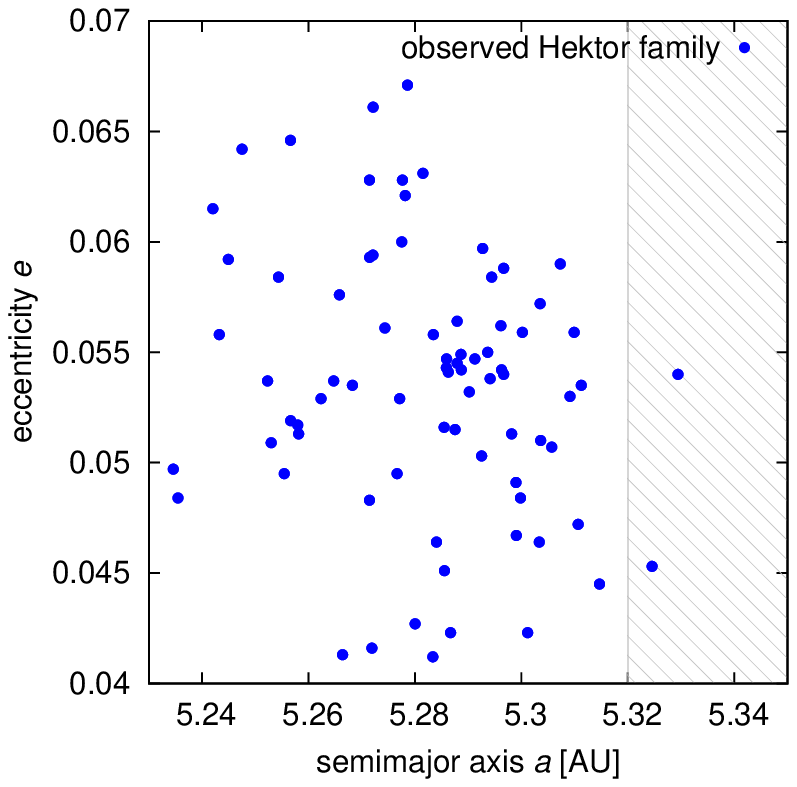} & 
\includegraphics[width=6cm]{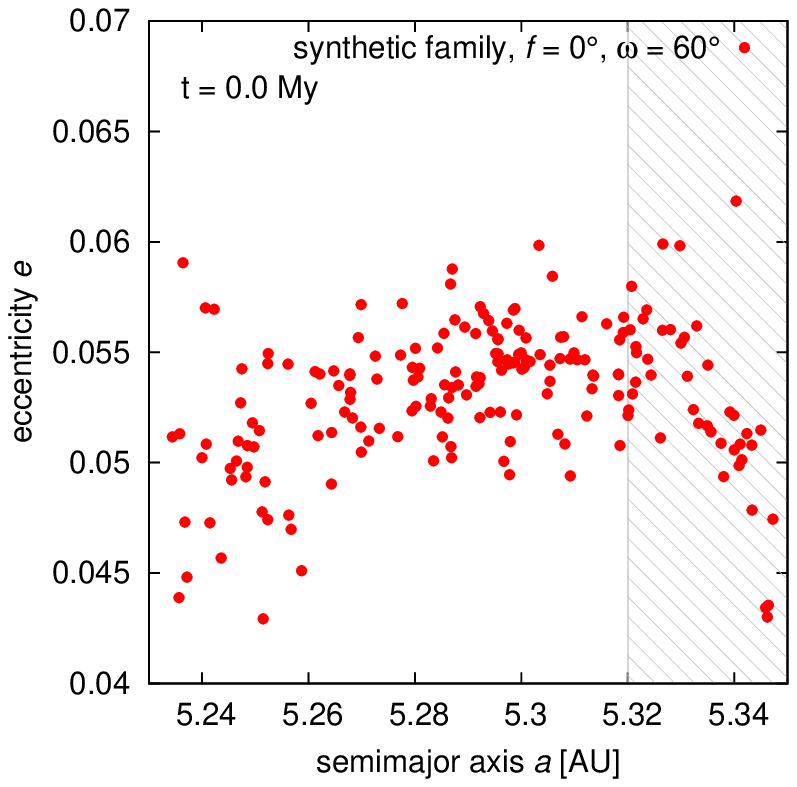} &
\includegraphics[width=6cm]{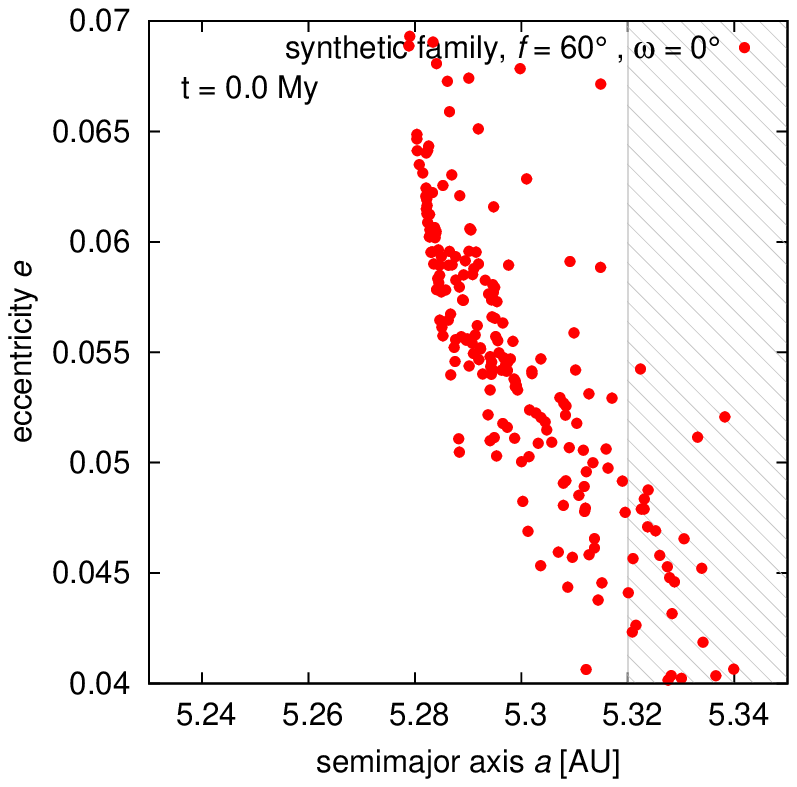}
\\
\includegraphics[width=6cm]{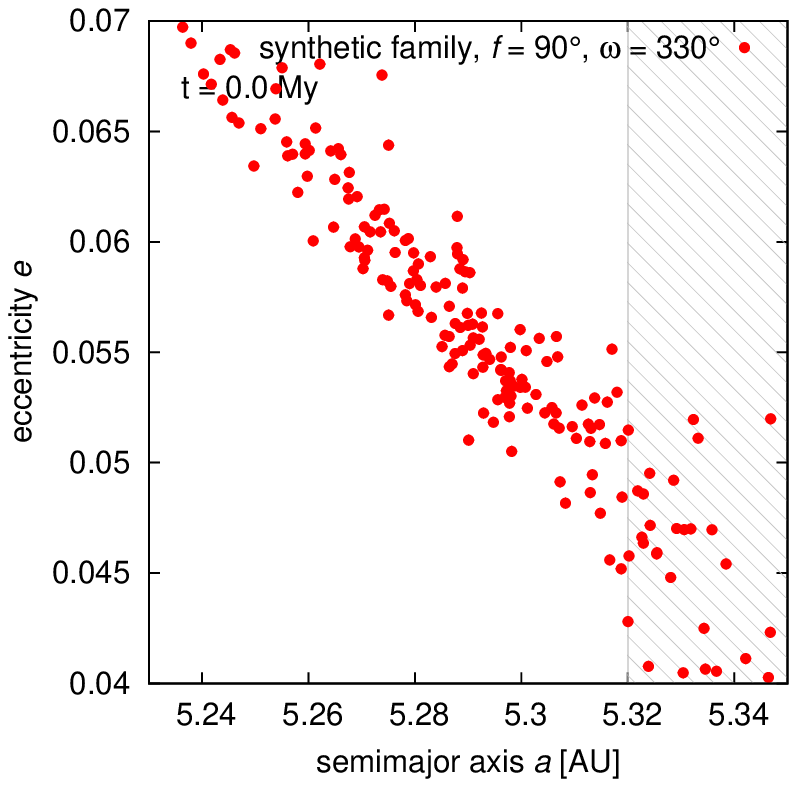} &
\includegraphics[width=6cm]{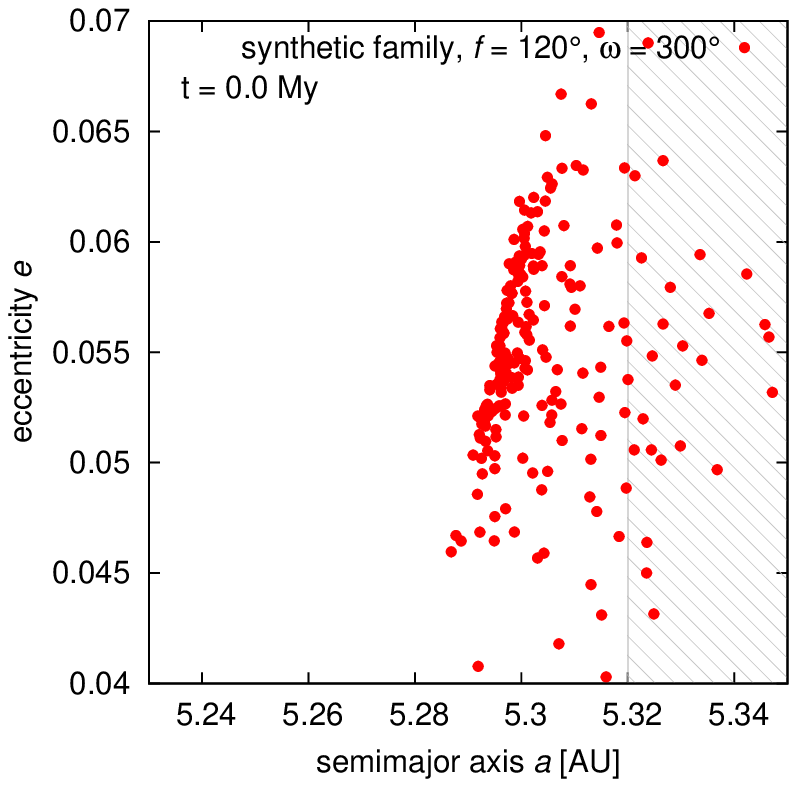} &
\includegraphics[width=6cm]{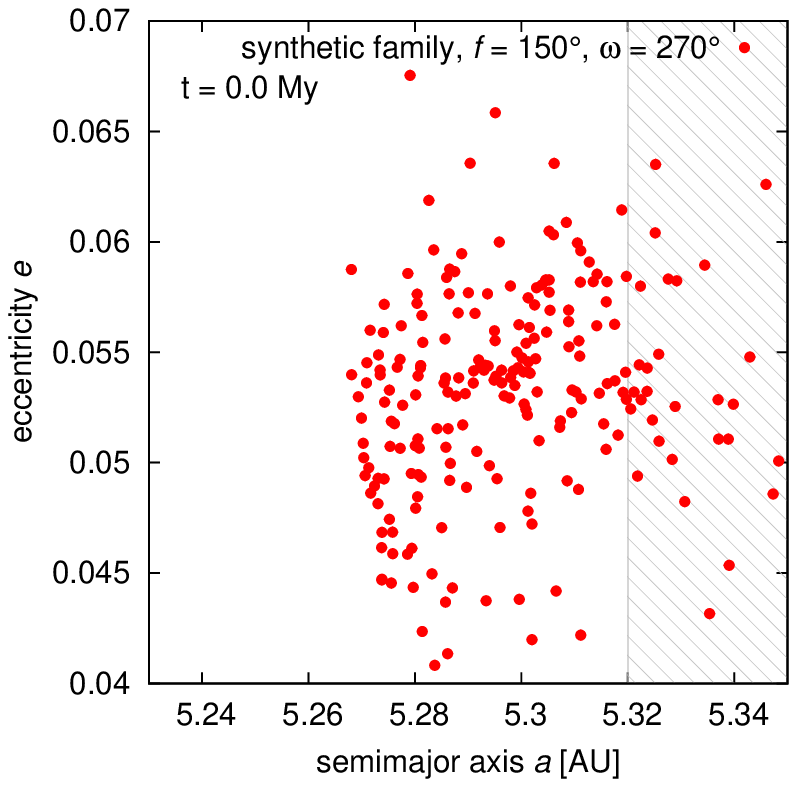} 
\\
\end{tabular}
\caption{Initial conditions for simulations of long-term evolution of synthetic 
families (red), compared to the observed Hektor family (blue) in the space of 
proper elements $(a_{\rm p},e_{\rm p})$. Each figure shows a different 
disruption geometry with different values of the true anomaly $f$ and 
the argument of pericentre $\omega$. Note the shaded area in the 
top left figure -- there are only two observed asteroids with proper semimajor 
axis $a_{\rm p} > 5.32\,\rm au$. This is due to the proximity to the 
border of the stable librating region. As there are many more synthetic 
asteroids in this region in all cases of initial distributions, we need to 
simulate a dynamical evolution of the family.}
\label{Hektor_sim_init}
\end{figure*}

\begin{figure*}
\centering
\renewcommand{\tabcolsep}{0pt}
\begin{tabular}{cc}
$(a_{\rm p},e_{\rm p})$  & $(a_{\rm p},I_{\rm p})$ \\
\includegraphics[width=9cm]{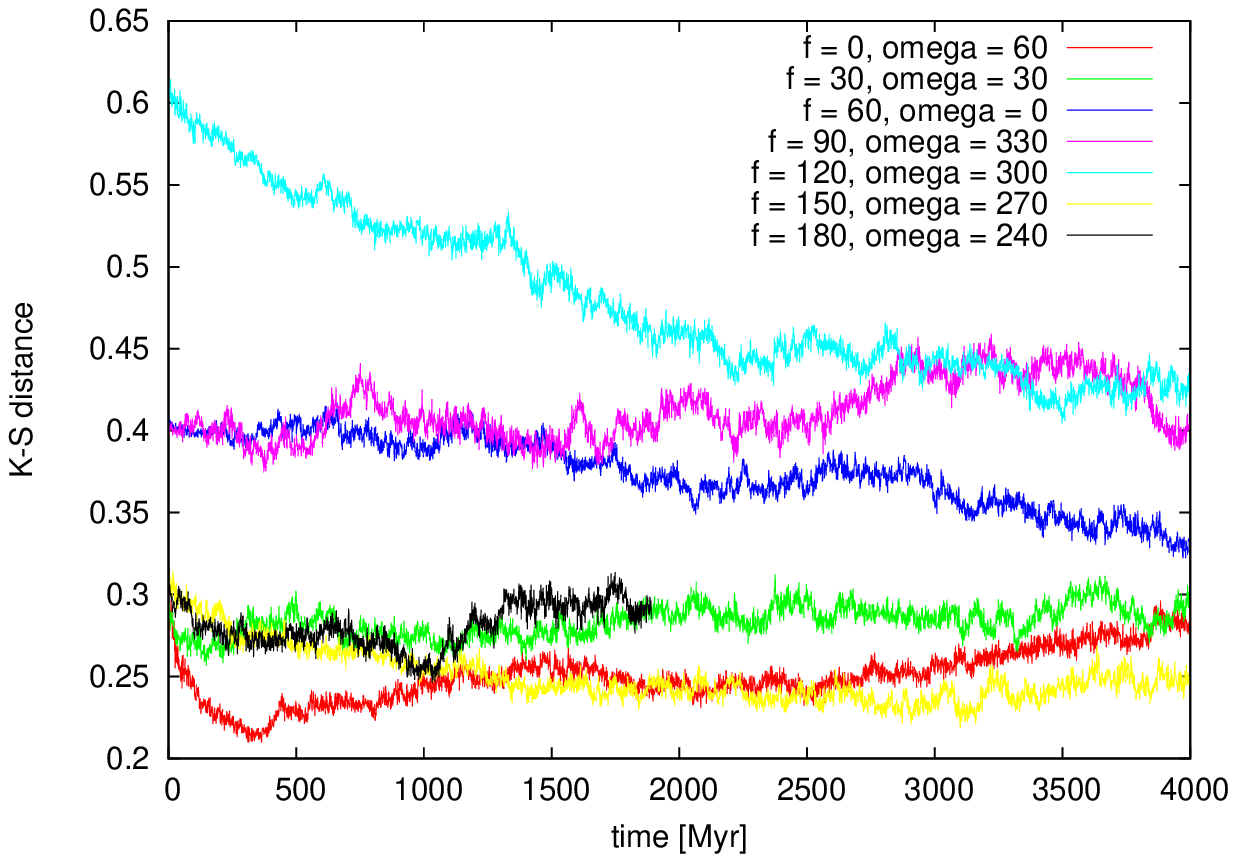} & 
\includegraphics[width=9cm]{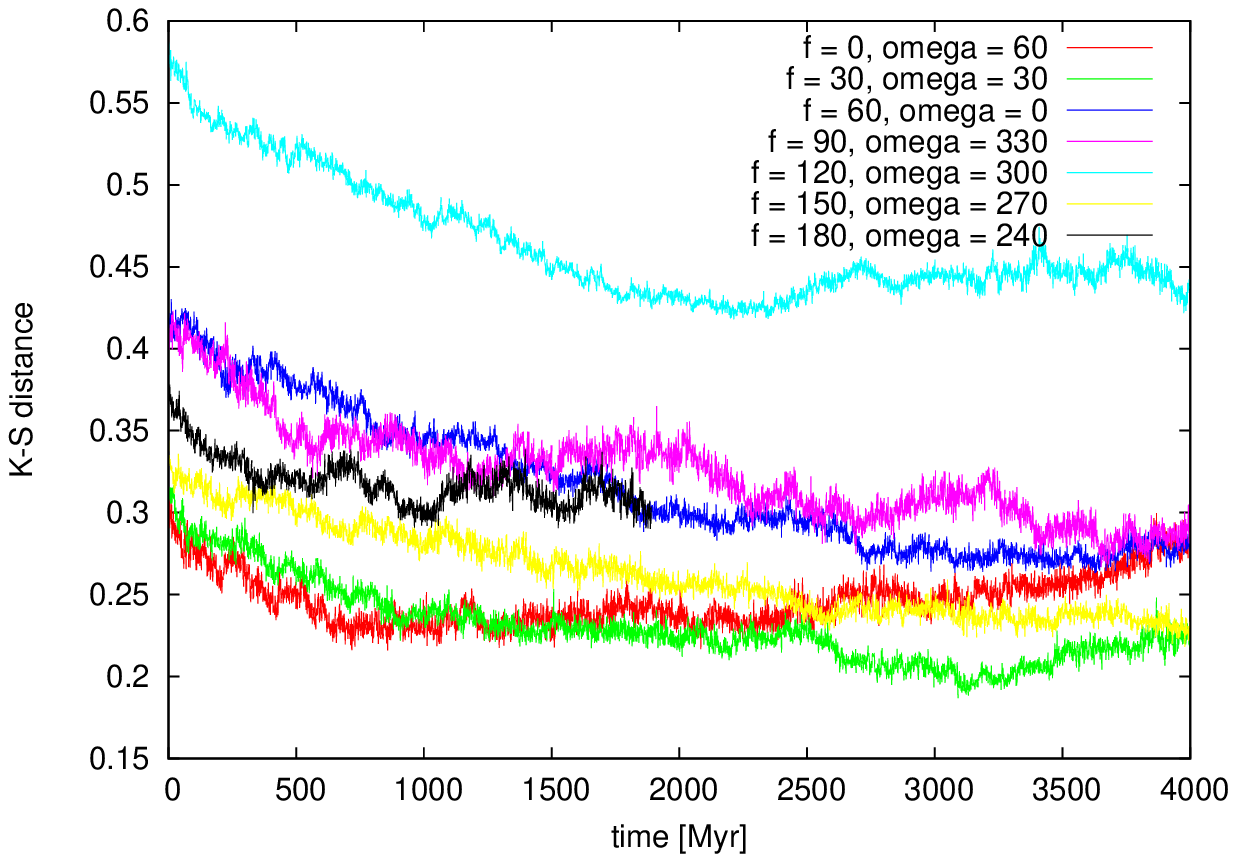} \\
\end{tabular}
\caption{Kolmogorov--Smirnov distance $D_{\rm KS}$ vs time for 7 different 
synthetic families compared with the observed Hektor family. Two-dimensional KS 
test was computed for the distributions of synthetic and observed 
families in the space of proper elements $(a_{\rm p},e_{\rm p})$ (left) and 
$(a_{\rm p},I_{\rm p})$ (right). The synthetic families were created assuming 
different impact geometries, namely the true anomaly 
$f=0^\circ,30^\circ,60^\circ, 90^\circ, 120^\circ, 150^\circ, 180^\circ $ and 
the argument of pericenter $\omega 
= 60^\circ, 30^\circ, 0^\circ, 330^\circ, 300^\circ, 270^\circ, 240^\circ$, 
which were combined so that the sum $f + \omega = 60^\circ$. The 
averaged distance $D_{\rm KS}$ changes in the course of dynamical evolution and 
we can see two minima: for $f=0^\circ$ and $\omega=60^\circ$ (red curve) it is
at about ($350 \pm 100$) Myr; for $f=150^\circ$ and 
$\omega=270^\circ$ (yellow curve) there is a flat minimum at ($2800 \pm 1500$) 
Myr. Since the red and yellow curves are overlapping in the range from 1800~Myr 
to 2500~Myr, we adopt the values of possible ages as 100 to 2500~Myr for the 
$f=0^\circ$ and $\omega=60^\circ$ geometry (red curve) and 1000 to 4000~Myr for 
the $f=150^\circ$ and $\omega=270^\circ$ geometry (yellow curve).}
\label{KS}
\end{figure*}

Initial positions of synthetic families members just after the disruption, 
compared to the observed Hektor family, are shown in Figure 
\ref{Hektor_sim_init}. To make a quantitative comparison of the distribution in 
the space of proper elements, we used a two-dimensional Kolmogorov--Smirnov 
test to compute KS distance of the synthetic family to the observed one with 
the output timestep of 1 Myr. The results for different initial geometries are
shown in Figure \ref{KS}.

\begin{figure*}
\centering
\begin{tabular}{cc}
\includegraphics[width=8.5cm]{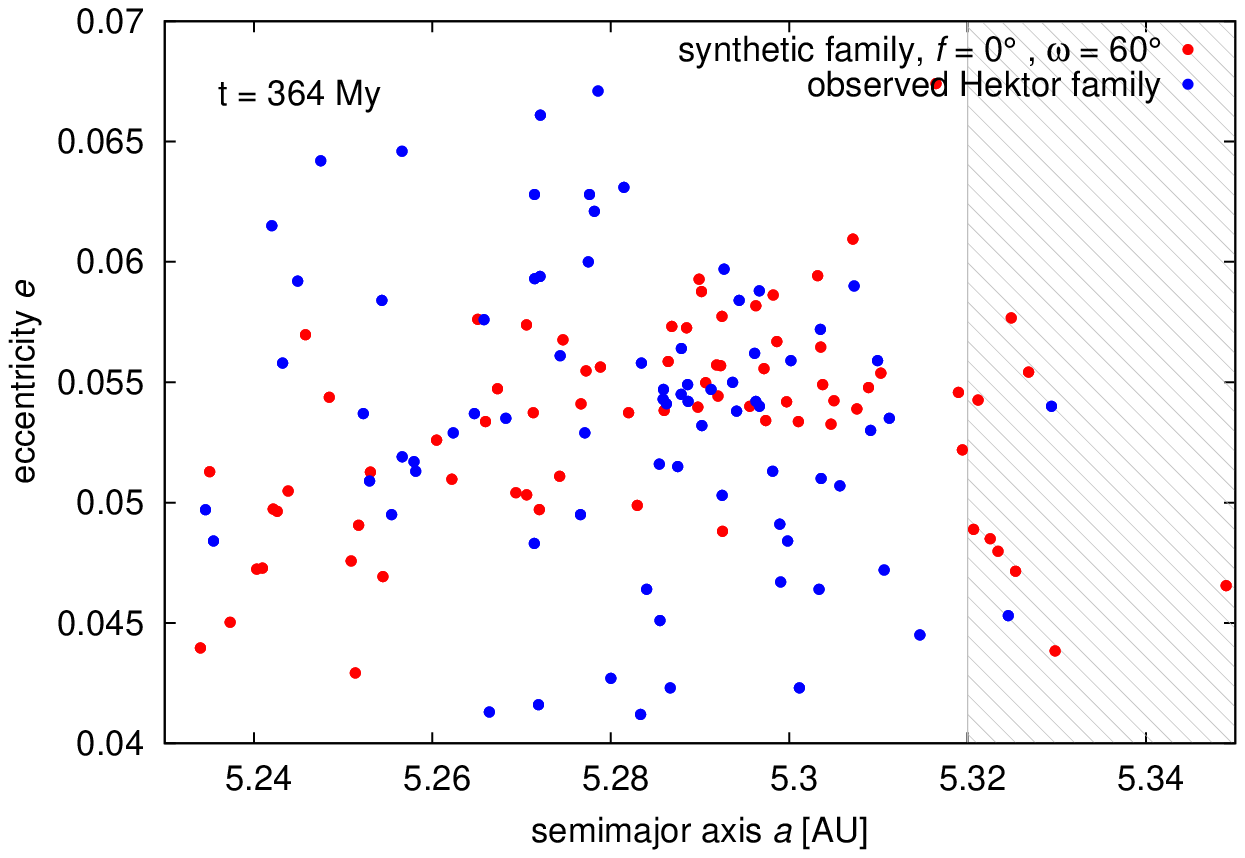} & 
\includegraphics[width=8.5cm]{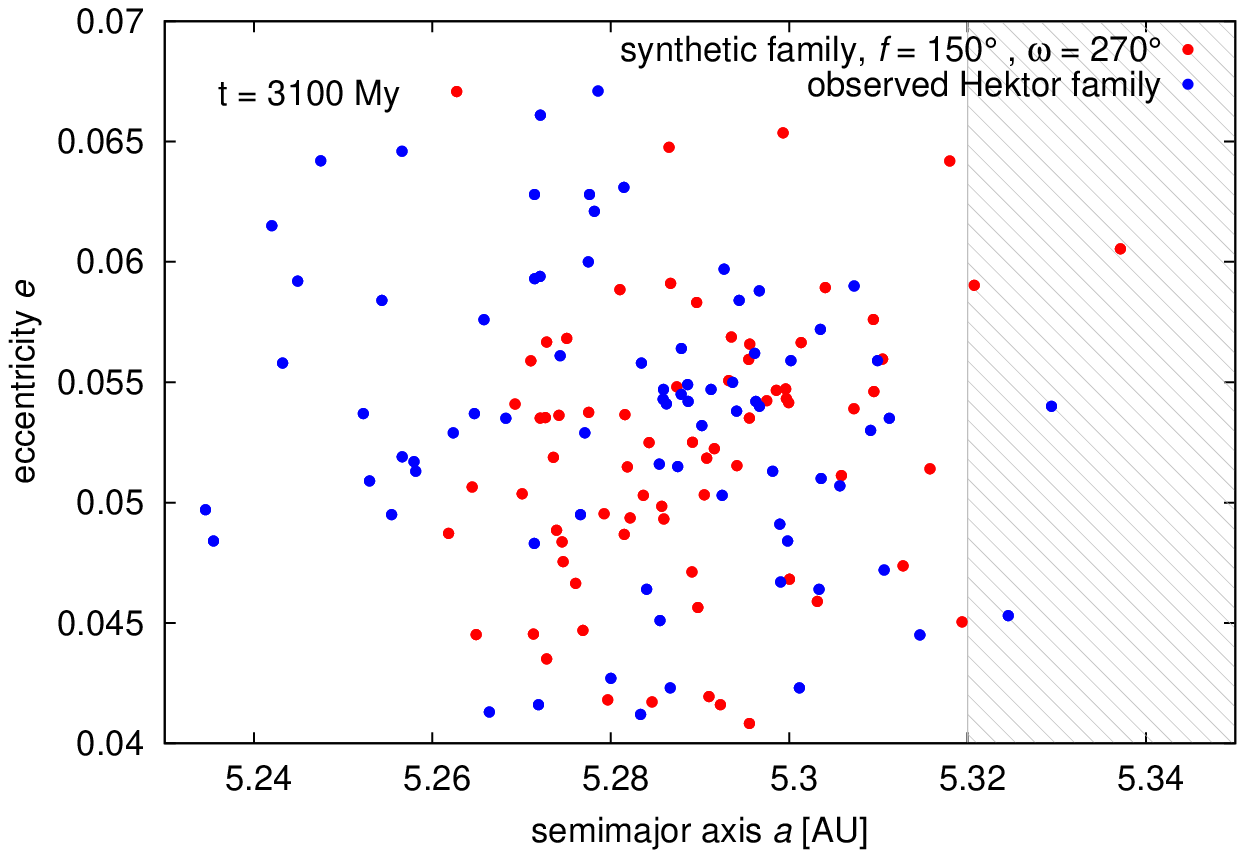} \\
\end{tabular}
\caption{Two evolved synthetic families in the 
space of proper elements $(a_{\rm p},e_{\rm p})$, which correspond to the 
minima of KS distance in Figure \ref{KS}.
Left picture shows the synthetic family (red) with $f=0^\circ$ and 
$\omega=60^\circ$ after 364 Myr of evolution in comparison with the observed 
Hektor family (blue). Right picture corresponds to the synthetic family with 
$f=150^\circ$ and $\omega=270^\circ$ after 3100 Myr of evolution. 
These two pictures differ in fine details, which cannot be accounted 
for in the KS statistics: i) the ``fibre-like'' structure of the 
relatively young family is still visible in the left picture; ii) there are 
many fewer synthetic bodies in the shaded area of the right picture $(a_{\rm 
p}>5.32\,\rm au)$ than on the left, which is closer to the observed reality.}
\label{Hektor_best_fit}
\end{figure*}

Our two best fits corresponding to the lowest KS distance are displayed in 
Figure \ref{Hektor_best_fit}. As we can see from the image of the whole Trojan 
L4 population, Hektor seems to be near the outskirts of the librating region 
(cf. Figure \ref{trojans_L4_arer_sizes}). In 
Figure~\ref{Hektor_sim_init}, we can note, that there are almost no observed 
asteroids in the shaded area with $a_{\rm p}> 5.32\,\rm{au}$, but we can see 
some synthetic family members in the left panel of Figure \ref{Hektor_best_fit} 
(initial geometry $f=0^\circ$, $\omega = 60^\circ$).

On the other hand, when we look at right panel of Figure \ref{Hektor_best_fit} 
(initial geometry $f=150^\circ$, $\omega = 270^\circ$), we can see that there 
are many fewer bodies in the proximity of the border of the stable librating 
region. One can also see the initial ``fibre-like'' structure is 
still visible on the left, but is almost dispersed on the right.

Hence, we conclude that the geometry at which the 
disruption occurred is rather $f=150^\circ$, $\omega = 270^\circ$ and the 
corresponding age is between 1 and 4 Gyr. The second but less likely 
possibility is that the disruption could have occured more recently (0.1 to 2.5 
Gyr) at $f=0^\circ$, $\omega = 60^\circ$.

\subsubsection{Parent body size from SPH simulations}\label{subsec:Durda_fit}
\begin{figure}
\centering
\includegraphics[width=7.0cm]{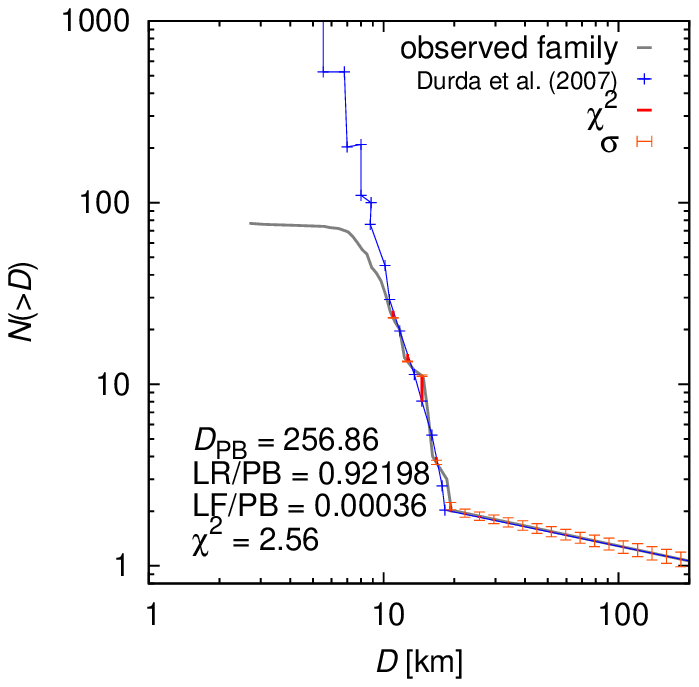}
\caption{Our best-fit size-frequency distribution of Hektor family  
by scaled SFDs from SPH simulations of Durda et al. (2007). 
In this particular case $D_{\rm PB(SPH)} = 257\,\rm km$, impactor 
diameter $D_{\rm imp} = 48\,\rm km$, impactor velocity $v_{\rm imp} = 
4\,\rm{km}\,\rm{s^{-1}}$ and impact angle $\varphi_{\rm imp} = 60^\circ$. 
However, other fits with similar pseudo-$\chi^2$ suggest the uncertainties are 
as follows: $\Delta D_{\rm PB(SPH)} = 10\,\rm km$,  $\Delta D_{\rm imp} = 
2\,\rm km$, $\Delta v_{\rm imp} = 1\,\rm{km}\,\rm{s^{-1}}$ and $\Delta 
\varphi_{\rm imp} = 15^\circ$. SFD shape seems to be more dependent on impact 
geometry than on impact velocity.}
\label{Hektor_SPH_fit}
\end{figure}

We tried to estimate the parent body size of Hektor family and other families 
by the method described in Durda et al. (2007). To this point, we calculated a 
pseudo$-\chi^2$ for the whole set of synthetic size-frequency distributions as 
given by the SPH simulations results (see Figure \ref{Hektor_SPH_fit}). 

Parent body sizes $D_{\rm PB(SPH)}$ and mass ratios of the largest fragment and 
parent body $M_{\rm{LF}}/M_{\rm{PB}}$ estimated by this method are listed in 
Table \ref{fams_prop}. The parent body size for Hektor family we derived 
from SPH simulations is $D_{\rm PB(SPH)} = (260 \pm 10)\,\rm km$, the impactor 
diameter $D_{\rm imp} = (24 \pm 2)\,\rm km$, the impactor velocity $v_{\rm imp} 
= (4 \pm 1)\,\rm{km}\,\rm{s^{-1}}$ and the impact angle $\varphi_{\rm imp} = 
(60^\circ \pm 15^\circ)$. We will use these values as initial 
conditions for simulations of collisional evolution below.

\subsection{1996\,RJ --- extremely compact family}
In our previous work, we mentioned a small cluster associated with asteroid 
(9799) 1996~RJ, which consisted of just 9~bodies. With the contemporary 
sample of resonant elements we can confirm that this cluster is indeed 
visible. It is composed of 18 bodies situated near the edge of the librating 
zone on high inclinations, within the ranges $I_{\rm p} \in \langle 31.38^\circ 
; 
32.27^\circ \rangle$ and $a_{\rm{p}}\in\langle 5.225\,;\,5.238\rangle$~au. As 
it is detached from the background in the space of proper elements, it remains 
isolated even at high cutoff velocity $v_{\rm{cutoff}}=160\, \rm{m\,s^{-1}}$.

Unfortunately, we have albedos measured by WISE for just 4 members of this 
family. These albedos are not much dispersed. They range from $p_{\rm{V}} = 
0.079 \pm 0.019$ to $p_{\rm{V}} = 0.109 \pm 0.029$ and, compared to the median 
albedo of the whole $L_4$ population $\widetilde {p_{\rm{V}}} = 0.072 \pm 
0.017$, they seem to be a bit brighter, but this statement is a bit 
inconclusive. 

\subsection{Arkesilaos}
This family is located on low inclinations $I_{\rm p} \in \langle 8.52^\circ ; 
9.20^\circ \rangle$, in the range of $a_{\rm{p}}\in\langle 
5.230\,;\,5.304\rangle$~au. It is clearly visible in the space of proper 
elements, although this area of $L_4$ cloud is very dense.

Still, it is difficult to find the largest remnant of the parent body, because 
this region is populated mainly by small asteroids with absolute magnitudes 
$H>12$. The only four asteroids with $H<12$ are (2148) Epeios with $H=10.7$, 
(19725)~$1999\,\rm WT_4$ with $H=10.7$, (38600)~$1999\,\rm XR_{213}$ with 
$H=11.7$ and (20961)~Arkesilaos with $H=11.8$. The only diameter derived from 
measured albedo is 
that of (2148) Epeios, which is $D = (39.02 \pm 0.65)\,\rm km$. Diameters of 
remaining bodies were calculated from their absolute magnitude assuming albedo 
$p_{\rm V} = 0.072$, which is the median of $L_4$ Trojans. Although (20961) 
Arkesilaos has the diameter only $D = (24 \pm 5)\,\rm km$, it is the only 
asteroid with $H<12$, for which the associated family has a reasonable number of 
members $N_{\rm memb}$ even for small values of the cutoff velocity $v_{\rm 
cutoff}$ (see Section \ref{sec:HCM}). As this is also the only larger body 
located near the center of the family in the space of proper elements, we 
treat (20961) Arkesilaos as the largest remnant of the parent body, whose 
diameter we estimate to be $D_{\rm PB(SPH)} \simeq 87\,\rm km $. Given that 
the mass ratio of the largest remnant and the parent body, as derived from 
SPH simulations of Durda et al. (2007), is $M_{\rm LR}/M_{\rm PB} \simeq 0.02$ 
only, it seems this family inevitably originated from a catastrophic disruption.

\subsection{Ennomos}
In our previous work, we reported a discovery of a possible family associated 
with asteroid (4709) Ennomos. With new data, we can still confirm that 
there is a significant cluster near this body, but when we take into account 
our ``$N_{\rm{memb}}(v_{\rm{cutoff}})$'' criterion described above, it turns 
out that the family is rather associated with asteroid (17492) Hippasos. It is 
a relatively numerous group composed of almost 100~bodies, situated near the 
border of the stable librating zone $L_5$ at high inclinations, ranging from 
$I_{\rm p} \in\langle 26.86^\circ;30.97^\circ \rangle$, and 
$a_{\rm{p}}\in\langle 5.225; 
5.338\rangle$~au.

\subsection{2001\,UV$_{209}$}
Using new data, we discovered a ``new'' family around asteroid (247341) 
$2001\,\rm UV_{209}$, which is the second and apparently the last observable 
family in our sample. Similar to the Ennomos family, it is 
located near the border of the $L_5$ zone on high inclinations $I_{\rm 
p} \in \langle 24.02^\circ;26.56^\circ \rangle$ and $a_{\rm{p}}\in\langle 
5.218;5.320\rangle$~au. This family has an exceptionally steep slope of the 
SFD, with $\gamma=-8.6 \pm 0.9$, which may indicate a recent collisional origin 
or a disruption at the boundary of the libration zone, which may be 
indeed size-selective as explained in Chrenko et al. (2015). 

%

\section{Collisional models of the Trojan population}\label{sec:coll_model}

In order to estimate the number of collisional families among $L_4$ Trojans, we 
performed a set of 100 simulations of the collisional evolution of Trojans with 
the Boulder code (Morbidelli et al., 2009) with the same initial 
conditions, but with different values of the random seed. 

\subsection{Initial conditions}
We set our initial conditions of the simulations such that 4 Gyr 
of collisional evolution leads to the observed cumulative SFD of $L_4$~Trojans 
(red curve in Figure \ref{fig:Boulder_SFD}). We constructed the 
initial synthetic SFD as three power laws with the slopes $\gamma_{\rm a} = 
-6.60$ in the size range from $D_1=117\,\rm{km}$ to $D_{\rm max}=250\,\rm{km}$, 
$\gamma_{\rm b} = -3.05$ from $D_2=25\,\rm{km}$ to $D_1$ and 
$\gamma_{\rm c} = -3.70$ from $D_{\rm min}=0.05\,\rm{km}$ 
to $D_2$. The synthetic initial population was normalized to 
contain $N_{\rm norm}=11$ asteroids with diameters $D\geq D_1$. 

To calculate the target strength $Q_D^*$, we used a parametric formula of 
Benz and Asphaug (1999):

\begin{equation}\label{Q*}
Q_D^*=Q_0 R_{\rm PB}^a + B\rho_{\rm bulk}R_{\rm PB}^b,
\end{equation}
where $R_{\rm PB}$ is the parent body radius in centimetres, $\rho_{\rm bulk}$ 
its bulk density, which we set to be $\rho_{\rm bulk} = 
1.3\,\,\rm{g}\,\rm{cm^{-3}}$ for synthetic Trojans (cf. Marchis et al., 2014). 
As of constants $a,\,b,\,B\,\rm{and}\,Q_0$ we used the values determined by 
Benz and Asphaug (1999) 
for ice at the impact velocity $v_{\rm imp}=3\,\rm km\,s^{-1}$, 
which are: $a=-0.39$, $b=1.26$, $B=1.2\,\,\rm erg\,cm^3\,g^{-2}$ and 
$Q_0=1.6\cdot10^7\,\rm erg\,g^{-1}$. 

In our model, we take into account only Trojan vs Trojan collisions, as 
the Trojan region is practically detached from the main belt. Anyway, 
main-belt asteroids with eccentricities large enough to reach the Trojan region 
are usually scattered by Jupiter on a time scale significantly shorter than the 
average time needed to collide with a relatively large Trojan asteroid. We 
thus assumed the values of collisional probability $P_{\rm i}=7.80\cdot 
10^{-18}\,\,\rm km^{-2}\,yr^{-1}$ and the impact velocity $v_{\rm 
imp}=4.66\,\,\rm km\,s^{-1}$ (Dell'Oro et al., 1998). Unfortunately, Benz and 
Asphaug (1999) do not provide parameters for ice at the impact velocities 
$v_{\rm imp} > 3\,\rm km\,s^{-1}$. 

We also ran several simulations with appropriate values 
for basalt at impact velocity $v_{\rm imp}=5\,\rm km\,s^{-1}$ ($a=-0.36$, 
$b=1.36$, $B=0.5\,\,\rm erg\,cm^3\,g^{-2}$ and $Q_0=9\cdot10^7\,\rm 
erg\,g^{-1}$).

Both models qualitatively exhibit the same evolution of SFD
and they give approximately the same total numbers of disruptions and 
craterings occured, but for basalt the model gives three times fewer 
\textit{observable} families originated by cratering than for 
ice. The results for the ice match the observation better, so we will further 
discuss the results for ice only.

\subsection{Long-term collisional evolution}
\begin{figure}
\centering
\includegraphics[width=8.5cm]{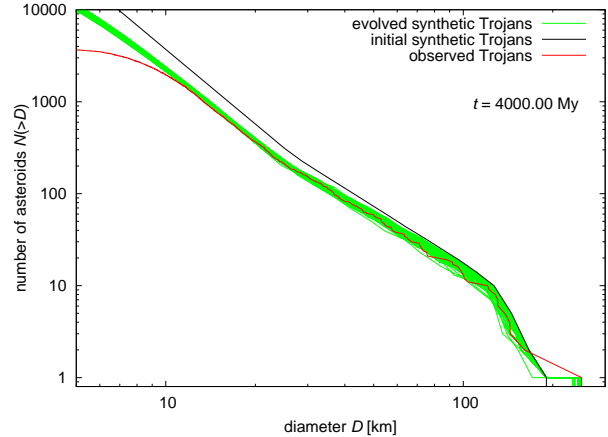}
\caption{Simulations of the collisional evolution of $L_4$ Trojans with the 
Boulder code (Morbidelli et al., 2009). Shown here is the initial cumulative 
SFD of a synthetic population (black) and the SFD of the observed one (red). 
Green are the final SFDs of 100 synthetic populations with the same initial SFD 
but with different random seeds, after 4 Gyr of a collisional evolution. The 
evolution of bodies larger than $D> 50\,\rm{km}$ is very slow, hence we can 
consider this part of the SFD as captured population.}
\label{fig:Boulder_SFD}
\end{figure}

The results of our simulations of the collisional evolution are shown in Figure 
\ref{fig:Boulder_SFD}. Our collisional model shows only little changes above 
$D>50\,\rm{km}$ over the last 3.85 Gyr (i.e. post-LHB phase only). Slopes of 
the initial synthetic population and the observed $L_4$ population 
differ by $\Delta \gamma < 0.1$ in the size range from 50~km to 100~km, while a 
relative decrease of the number of asteroids after 3.85 Gyr of collisional 
evolution is 
only about $12\,\%$ in the same size range. Hence, we can consider this part of 
the Trojan population as a representative sample of the source population, 
which 
is not much affected by collisional evolution. Therefore, these Trojans provide 
very useful information about the source population, from which they were 
captured (as modeled in Nesvorn\'y et al., 2013).

\begin{figure*}
\centering
\renewcommand{\tabcolsep}{0pt}
\begin{tabular}{cc}
\includegraphics[width=9cm]{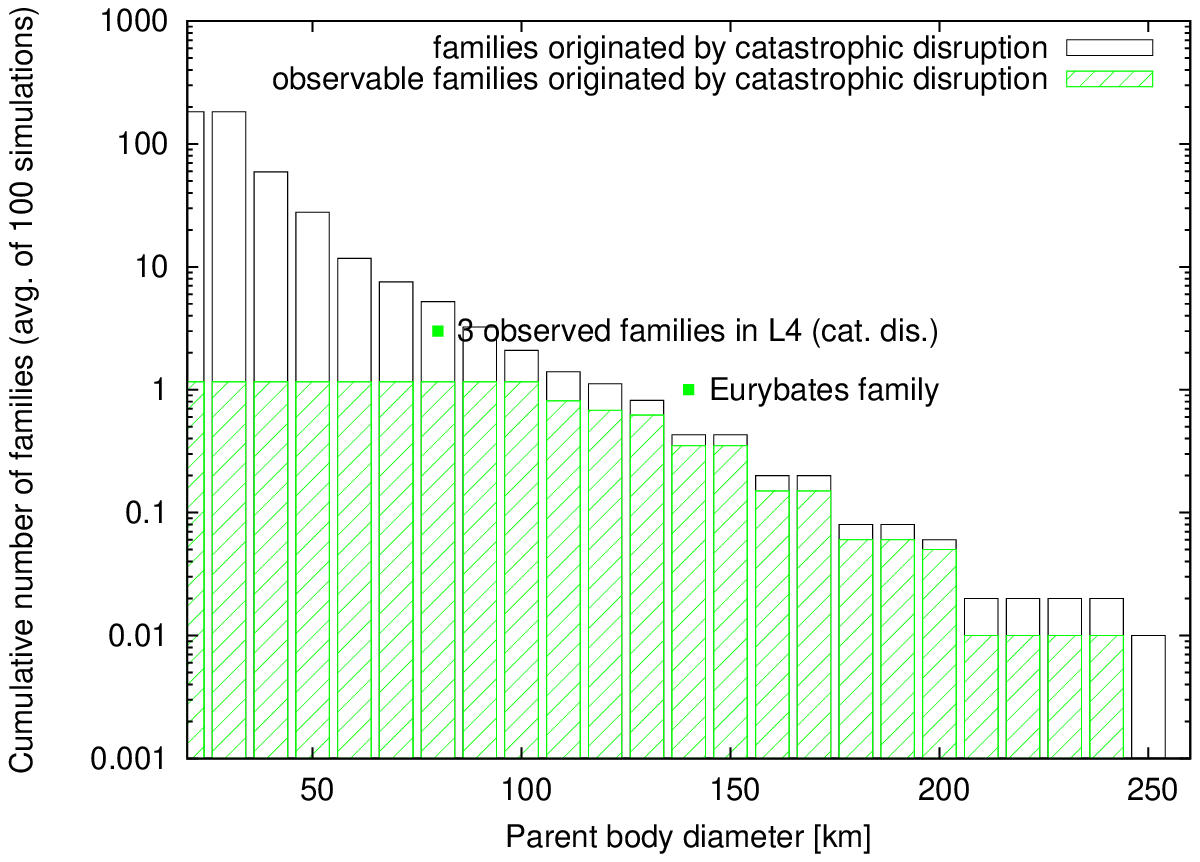} & 
\includegraphics[width=9cm]{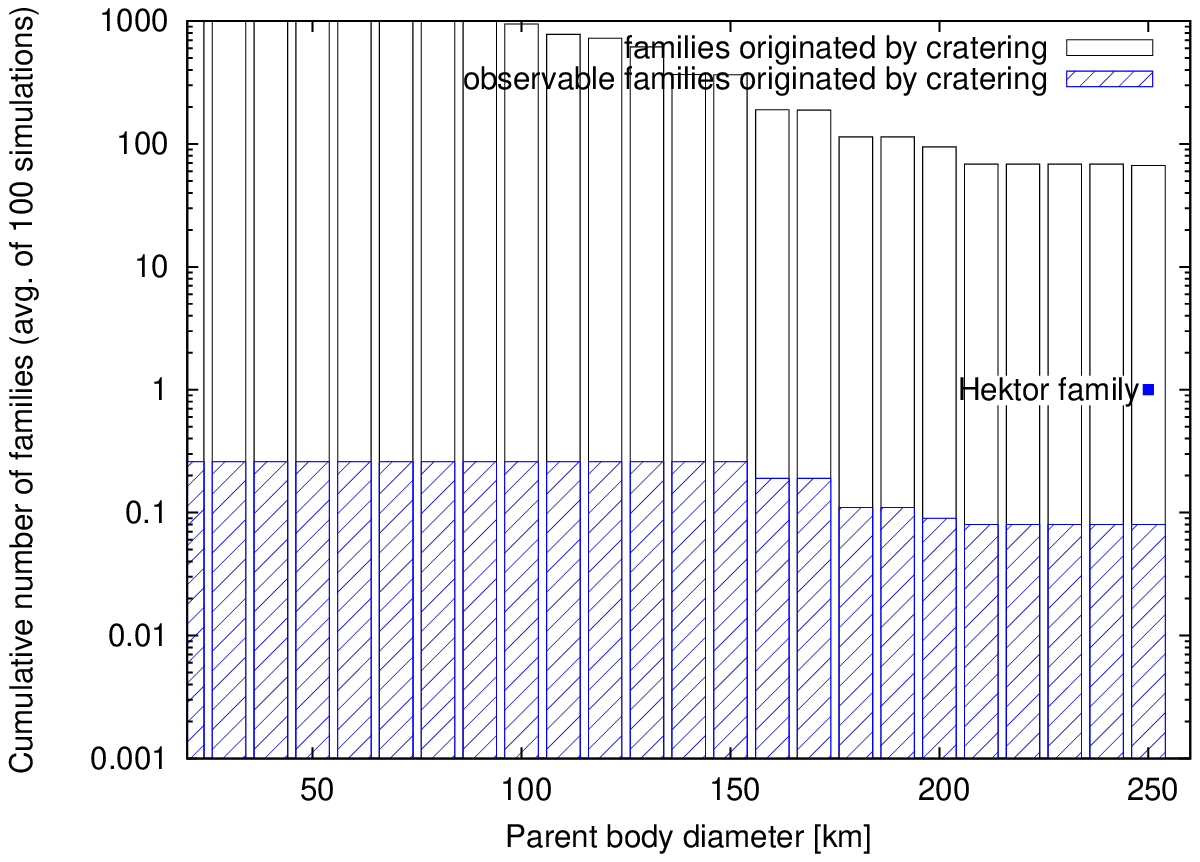} \\
\end{tabular}
\caption{The dependence of the cumulative number (an average over 
100 simulations) of catastrophic disruptions among Trojans (left panel) and 
cratering events 
(right panel) on the target diameter $D_{\rm PB}$ (black boxes), and a subset 
of those Trojan families, which should be detected in contemporary 
observational data, 
i.e. with the number of fragments $N(D>10\,\rm{km})>10$ (green boxes for 
disruptions and blue boxes for craterings). In other words, colour boxes 
represent simulated detections of families based on the expected 
effectiveness of our detection methods. This is 
the reason, why the cumulative number of the observable families does not 
strictly increase with the decreasing parent body size, but is rather constant 
under the limit of about 95~km in the case of catastrophic disruptions and 
145~km in the case of craterings. There are also observed 
families marked for a comparison. Three of the four observed families in $L_4$ 
cloud originated by catastrophic disruption, while only one (Hektor) originated 
by cratering event (cf. Table \ref{fams_list}).}
\label{fig:Boulder_families}
\end{figure*}

\subsection{An estimate of the number of observable families}
From our set of simulations, we also obtained the number of collisions leading 
to collisional families among $L_4$ Trojans, namely catastrophic 
disruptions, where the mass ratio of the largest remnant and the parent body 
$M_{\rm{LR}}/M_{\rm{PB}} < 0.5$, and cratering events, where 
$M_{\rm{LR}}/M_{\rm{PB}} > 0.5$. As one can verify in Figure 
\ref{fig:Boulder_families}, these numbers are dependent on the diameter of the 
parent body $D_{\rm PB}$.  

However, not all of these collisions produce families which are in fact 
observable (detectable). There are generally two possible obstacles in 
the detection of a family in the space of proper elements: i) 
somewhat more concentrated background population, due to which our detection 
methods (both ``randombox'' and HCM, see Chapter \ref{sec:groups}) may fail, if 
the number of observed fragments is too low in comparison with the background, 
and ii) an observational incompleteness, which means that in the case of 
Trojans, a substantial part of fragments with sizes $D<10\,\rm km$ is still 
unknown, what again reduces a chance of a family detection.

For these reasons, we constructed a criterion of observability that a 
synthetic family must fulfill in order to be detectable in the current 
conditions (i.e. we simulated a detection of synthetic families by the 
same methods we used to detect the real ones). The simplest criterion could be 
that a family must contain at least $N_{\rm min} = 10$ fragments with diameter 
$D\geq 10\,\rm{km}$. 

Within 100 simulations, there were 93 catastrophic disruptions of bodies 
with diameters $D_{\rm{PB}}>100\,\rm{km}$, but only 50 of them produced more 
than 10 fragments with $D\geq10\,\rm{km}$, see Figure 
\ref{fig:Boulder_families}. Hence, the probability that we would observe a 
collisional family originated by a catastrophic disruption of a parent body 
with 
$D_{\rm{PB}}>100\,\rm{km}$ is \textit{only} 0.50, which matches the 
observations 
(namely Eurybates family with $D_{\rm PB(SPH)}\simeq155\,\,\rm km$, see Table 
\ref{fams_prop}). This value is also roughly consistent with our previous 
estimate based on the stationary model (Bro\v{z} and Rozehnal, 2011), which 
gives the value 0.32 with new observational data. 

As one can also see in Figure \ref{fig:Boulder_families}, the number of 
cratering events is about one to two orders higher than the number of 
catastrophic disruptions, however, they do not produce enough fragments larger 
than $D\geq10\,\rm{km}$. For the parent body size $D_{\rm{PB}}>100\,\rm{km}$ 
there occurred almost 45,000 cratering events within 100 simulations which 
produced the largest fragment with $D_{\rm LF}\geq1\,\rm km$, but only 10 of 
them fulfill our criterion of observability. Hence, the probability that we can 
observe a family originated by a cratering of a parent body with 
$D_{\rm{PB}}>100\,\rm{km}$ is only 0.10, at least with contemporary data. From 
a statistical point of view, this can actually correspond to the Hektor family. 

As we have already demonstrated in Bro\v{z} and Rozehnal (2011), the number of 
families is not significantly affected by chaotic diffusion 
or by a ballistic transport outside the libration zone.


\section{SPH simulations of Hektor family}\label{sec:SPH}

As we have already mentioned in Section \ref{subsec:Hektor}, (624)~Hektor 
is very interesting Trojan asteroid with possibly bilobed shape and a small 
moon. Diameters of (624) Hektor stated in Marchis et al. (2014) are as 
follows: equivalent diameter $D_{\rm eq}=(250\pm26)\,\,\rm km$ for a  
convex model, the individual diameters of the lobes $D_{\rm 
A}=(220\pm22)\,\,\rm km$, $D_{\rm B}=(183\pm18)\,\,\rm km$ for a bilobed 
version. Estimated parameters of the moon are: the diameter $D_{\rm 
m}=(12\pm3)\,\,\rm km$, the semimajor axis $a_{\rm 
m}=(623\pm10)\,\,\rm km$,
the eccentricity $e_{\rm m}=(0.31\pm0.03)$ and 
the inclination (with respect to the primary equator)
$I_{\rm m}=(50\pm1)^\circ$. 

As we associate (624) Hektor with the collisional family, we would like to 
know, how the properties of the family are influenced by the shape of target 
body. We therefore performed a series of SPH simulations aiming to explain the 
origin of the Hektor family, for both cases of convex and bilobed shape of its 
parent body. 

\subsection{Methods and initial conditions}

We simulated a collisional disruption using the smoothed-particle hydrodynamic 
code SPH5 (Benz and Asphaug, 1994). We performed two sets of simulations. 
In the first one, we simulated an impact on a single spherical asteroid. In 
the second, on a bilobed asteroid represented by two spheres positioned next to 
each other. The two touching spheres have a narrow interface,
so that the SPH quantities do not easily propagate
between them. In this setup, we are likely to see
differences between sinlge/bilobed cases as clearly
as possible. 

As for the main input parameters (target/impactor sizes, the impact velocity 
and the impact angle) we took the parameters of our best-fit SFDs, obtained 
by Durda et al. (2007) scaling method, see Section \ref{subsec:Durda_fit} and 
Figure \ref{Hektor_SPH_fit}. 

To simulate a collision between the parent body and the impactor 
we performed a limited set of simulations: 
i) a single spherical basalt target with diameter $D_{\rm PB} = 260\,\,\rm km$ 
vs a basalt impactor with diameter $D_{\rm imp} = 48\,\,\rm km$;
ii) the single basalt target $D_{\rm PB} = 260\,\,\rm km$ vs an ice 
impactor (a mixture of ice and 30\,\% of silicates) with $D_{\rm imp} 
= 64\,\,\rm km$ (impactor diameter was scaled to get the same kinetic energy); 
iii) a bilobed basalt target approximated by two spheres with diameters 
$D_{\rm PB} = 200\,\,\rm km$ each (the total mass is approximately the same) vs 
a basalt impactor with $D_{\rm imp} = 48\,\,\rm km$;
iv) a single spherical ice target $D_{\rm PB} = 260\,\,\rm km$ vs an ice 
impactor $D_{\rm imp} = 38\,\,\rm km$ (impactor diameter was 
scaled to get the same ratio of the specific kinetic energy $Q$ to the target 
strength $Q^*_{\rm D}$).

The integration was controlled by the Courant number $C = 1.0$, a typical 
time step thus was $\Delta t \simeq 10^{-5}\,{\rm s}$, and the time span was 
$t_{\rm stop} = 100\,{\rm s}$. The Courant condition was the same in different 
materials, using always the maximum sound speed $c_{\rm s}$ among
all SPH particles, as usually. 

We used $N_{\rm SPH, st}=10^5$~SPH particles for 
the single spherical target and $N_{\rm SPH, bt}=2\cdot 10^5$ for the bilobed 
one. For impactor $N_{\rm SPH, i}=10^3$ SPH particles. We assumed the 
Tillotson equation of state (Tillotson, 1962) and material properties, which 
are listed in Table \ref{tab:materials}. 

\begin{table}\vspace{-5mm}\centering
\caption{Material constants used in our SPH simulations for basalt and 
silicated ice (30\,\% of silicates). 
Listed here are:
the zero-pressure density $\rho_0$,
bulk modulus $A$,
non-linear compressive term $B$,
sublimation energy $E_0$,
Tillotson parameters $a$, $b$, $\alpha$ and $\beta$, 
specific energy of incipient vaporization $E_{\rm iv}$, 
complete vaporization $E_{\rm cv}$, 
shear modulus $\mu$, 
plastic yielding $Y$,
melt energy $E_{\rm melt}$
and Weibull fracture parameters $k$ and $m$. 
Values we used for silicated ice are identical to those of pure ice, except 
density $\rho_0$, bulk modulus $A$ and Weibull parameters $k$ and $m$.
All values were adopted from Benz and Asphaug (1999).
} 
\small
\renewcommand{\tabcolsep}{6pt}
\begin{tabular}{c|c|c|c|}	
\hline\
quantity & basalt & silicated ice & unit\\
\hline\hline
$\rho_0$ & 2.7 & 1.1 & ${\rm g}\,{\rm cm}^{-3}$ \\
$A$ & $2.67\cdot10^{11}$ & $8.44\cdot10^{10}$ & ${\rm erg}\,{\rm cm}^{-3}$ \\
$B$ & $2.67\cdot10^{11}$ & $1.33\cdot10^{11}$ & ${\rm erg}\,{\rm cm}^{-3}$ \\
$E_0$ & $4.87\cdot10^{12}$ & $1.00\cdot10^{11}$ & ${\rm erg}\,{\rm g}^{-1}$ \\
$a$ & 0.5 & 0.3 & -- \\
$b$ & 1.5 & 0.1 & -- \\
$\alpha$ & 5.0 & 10.0 & -- \\
$\beta$ & 5.0 & 5.0 & -- \\
$E_{\rm iv}$ & $4.72\cdot10^{10}$ & $7.73\cdot10^{9}$ & ${\rm erg}\,{\rm 
g}^{-1}$ \\
$E_{\rm cv}$ & $1.82\cdot10^{11}$ & $3.04\cdot10^{10}$ & ${\rm erg}\,{\rm 
g}^{-1}$ \\
$\mu$ & $2.27\cdot10^{11}$ & $2.80\cdot10^{10}$ & ${\rm erg}\,{\rm cm}^{-3}$ \\
$Y$ & $3.5\cdot10^{10}$ & $1.0\cdot10^{10}$ & ${\rm erg}\,{\rm g}^{-1}$ \\
$E_{\rm melt}$ & $3.4\cdot10^{10}$ & $7.0\cdot10^{9}$ & ${\rm erg}\,{\rm 
g}^{-1}$ \\
$k$ & $4.0\cdot10^{29}$ & $5.6\cdot10^{38}$ & ${\rm cm}^{-3}$ \\
$m$ & 9.0 & 9.4 & -- \\
\hline
\vspace{2mm} 
\end{tabular}
\label{tab:materials}
\end{table}

We terminated SPH simulations after 100 s from the impact. This time 
interval is needed to establish a velocity field of fragments and to complete 
the fragmentation.
Then we handed the output of the SPH simulation as initial 
conditions to the N--body gravitational code Pkdgrav (Richardson et al., 2000), 
a parallel tree code used to simulate a gravitational reaccumulation of 
fragments. Unlike Durda et al. (2007), who calculated radii of fragments $R$ 
from the smoothing length $h$ as $R=h/3$, we calculated fragments radii from 
their masses $m$ and densities $\rho$ as $R = (m/(4\pi\rho))^{1/3}$.

We ran Pkdgrav with the time step $\Delta t = 5.0\,\,\rm s$ and we terminated 
this simulation after $t_{\rm evol} = 3$~days of evolution. To ensure 
this is sufficiently long, we also ran several simulations with $t_{\rm evol} = 
5$~days, but we had seen no significant differences between final results.

We used the nominal value for the tree opening angle,
${\rm d}\theta = 0.5\,{\rm rad}$, even though for the
evolution of eventual moons it would be worth to use even
smaller value, e.g. ${\rm d}\theta = 0.2\,{\rm rad}$. 

\subsection{Resulting size-frequency distributions}

\begin{figure}
\centering
\includegraphics[width=8.5cm]{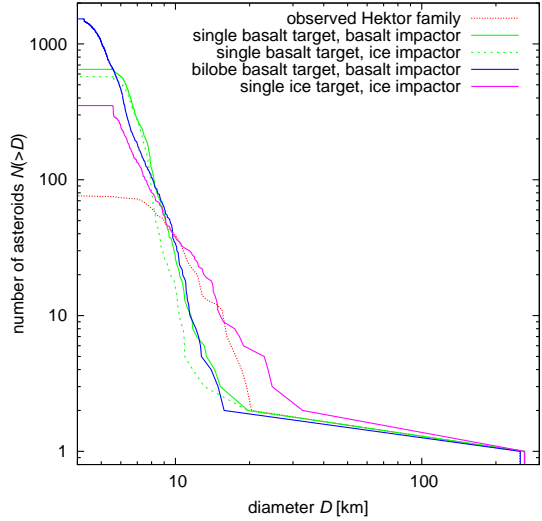}
\caption{A comparison of size-frequency distributions of the 
observed Hektor 
family (red dotted) and SFDs of synthetic families created by different SPH 
simulations, always assuming the impactor velocity $v_{\rm imp} = 4\,\,\rm 
km\,s^{-1}$ and the impact angle $\varphi_{\rm imp}~=~60^\circ$. For a single 
spherical target (green lines) we assumed the diameter $D_{\rm PB} = 260\,\,\rm 
km$, for a bilobe target (blue line) we approximated the lobes as spheres with 
diameters $D_{\rm PB} = 200\,\,\rm km$ each. The impactor size was assumed 
to be $D_{\rm imp} = 48\,\,\rm km$ in the case of 
basalt, $D_{\rm imp} = 64\,\,\rm km$ in the case of silicate ice impacting on 
basalt target (scaled to the same $E_{\rm imp}$) and $D_{\rm imp} = 38\,\,\rm 
km$ in the case of silicate ice impacting on ice target (scaled to the same 
$Q/Q^*_{\rm D}$). Fragments of the impactor were purposely removed from 
this plot, as they do not remain in the libration zone for our particular 
impact orbital geometry.}
\label{fig:SPH_SFD}
\end{figure}

From the output of our simulations we constructed size-frequency distributions  
of synthetic families, which we compare to the observed one, as demonstrated in
Figure \ref{fig:SPH_SFD}. As one can see, there are only minor 
differences between SFDs of families created by the impacts on the single and 
bilobed target, except the number of fragments with diameter $D<5\,\rm km$, 
but this is mostly due to different numbers of SPH particles. However, there 
are differences between ice and basalt targets. Basalt targets provide 
generally steeper SFDs with smaller largest remnants than the ice target.  

To make the comparison of these synthetic initial SFDs to each 
other more realistic, we removed the fragments of the impactor from our 
synthetic families. This is because fragments of the impactor often do not 
remain in the libration zone. Note that this procedure does not subtitute for a 
full simulation of further evolution; it serves just for a quick comparison of 
the SFDs.

To match the observed SFD of the Hektor family more accurately, we should 
perform a much larger set of simulations with different sizes of projectiles 
and also different compositions (mixtures of ice and basalt). However, material 
parameters of these mixtures are generally not known. Regarding the material 
constants of pure ice, we have them for the impact velocity $v_{\rm imp} = 
3\,\rm km\,s^{-1}$ only (Benz and Aspaugh, 1999). 
There are also some differences between SFDs of single and bilobe 
targets, so we should perform these simulations for each target geometry.  
However, we postpone these detailed simulations for future work; in this 
work we further analyse results of simulations with basalt targets and we 
focus on the evolution of the SFDs.

\begin{figure}
\centering
\includegraphics[width=8.5cm]{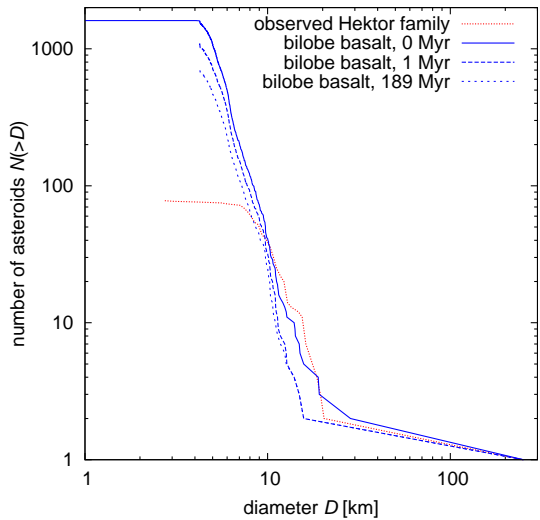}
\caption{A simulation of evolution of the SFD of a synthetic Hektor family due 
to a ballistic transport and chaotic diffusion. One can see here a rapid change 
of SFD within the first 1 Myr after the breakup as the fragments of the 
impactor leaved the libration zone in our impact geometry. This ballistic 
transport resulted in a reduction of the number of particularly larger bodies 
in our case. Further evolution due to the chaotic diffusion seems to cause the 
reduction of mostly smaller bodies. Note that the initial SFD (0 
Myr) contains some fragments of the impactor, so the blue solid curve looks 
different than the curve in Figure \ref{fig:SPH_SFD}, where the fragments of the 
impactor were removed.}
\label{fig:SPH_SFD_evol}
\end{figure}

It should be emphasized that the SFDs presented here correspond to very 
young synthetic families, hence they are not affected by any dynamical and 
collisional evolution yet. To reveal possible trends of the evolution by 
a ballistic transport and chaotic diffusion, we prepared initial conditions for 
the SWIFT integrator, similarly as described in Section \ref{sub:dyn_evol}, let 
the simulation run and monitored the corresponding evolution of the SFD. The 
results can be seen in Figure~\ref{fig:SPH_SFD_evol}. The biggest difference 
between $t=0$ and $t=1\,\,\rm Myr$ is caused by a ballistic transport outside 
the libration zone --- fragments (especially of the impactor) 
missing from the SFD at $t=1\,\,\rm Myr$ were perturbed too much to
remain in the libration zone, at least for a given impact geometry. We 
actually tested two impact geometries: in the direction tangential and 
perpendicular to the orbit.

This may be important for the method we used in Section \ref{subsec:Durda_fit} 
to derive a preliminary parent body size and other properties of the family. The 
SFDs obtained by Durda et al. (2007) were directly compared in their work to the 
main-belt families, however, there is a part of fragments among Trojans (in our 
case even the largest ones, see Figure \ref{fig:SPH_SFD_evol}), which cannot be 
seen in the space of resonant elements, because they do not belong to Trojans 
any more. Fortunately, values of pseudo-$\chi^2$ we computed in Section 
\ref{subsec:Durda_fit} depend rather weakly on the distribution of a few 
largest bodies. Even so, we plan to analyze SFDs of synthetic families more 
carefully in future works.

\begin{figure}
\centering
\includegraphics[width=8.5cm]{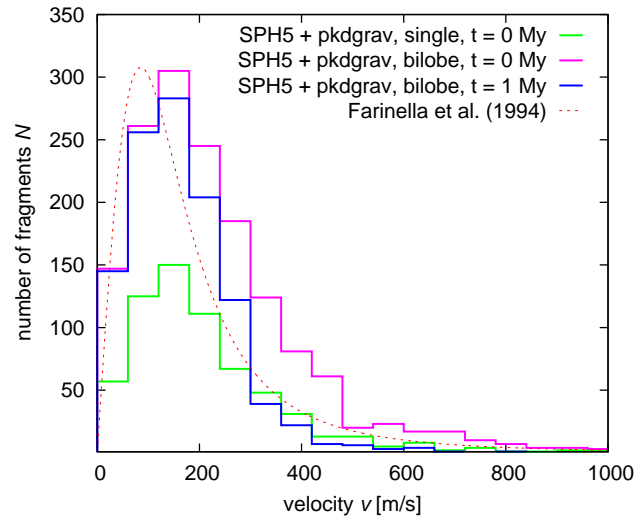}
\caption{Velocity distributions of fragments originated in various SPH 
simulations (green, violet) in comparison with the model of Farinella et al. 
(1994) we used in our $N$-body simulations of isotropic disruption and 
dynamical evolution (see Section \ref{sub:dyn_evol}). Shown here is also the 
distribution of velocities after 1 Myr of evolution, i.e. of fragments that 
remained in libration zones.}
\label{fig:SPH_v_evol}
\end{figure}

\subsection{Resulting velocity fields}\label{subsec:velocity_fields}

In our $N$-body simulations, we used the model of isotropic disruption 
(Farinella et al., 1994). As we compared the synthetic family with the observed 
one (see Section \ref{sub:dyn_evol}), we simulated only the evolution of bodies 
with relatively low ejection velocities ($v<200\,\rm m\,s^{-1}$), 
because the observed family is confined by the cutoff velocity $v_{\rm 
cutoff} = 110\,\rm m\,s^{-1}$. Very small fragments with higher velocities 
may be still hidden in the background.

Here, we compare Farinella's model to the velocity fields of fragments from SPH 
simulations, see Figure 
\ref{fig:SPH_v_evol}. We realized that Farinella's model is not offset 
substantially with respect to other velocity histograms, especially
at lower velocities, $v < 200\,{\rm m}\,{\rm s}^{-1}$.  On the other side, 
there remained some fragments of the impactor with velocities $v>2\,\rm 
km\,s^{-1}$
in our SPH simulations, which are not produced in the isotropic model. It does 
not affect a comparison of the synthetic and observed families in the space of 
proper elements, as these high-velocity fragments leaved the Trojan region in 
our case, but it does affect the SFD of the synthetic family. As a consequence, 
one should always analyse SFDs and velocity fields together. 

We also simulated a further evolution of the velocity field. After just 1 
Myr of evolution, there remained no bodies with $v>1.5\,\rm km\,s^{-1}$ in our 
impact geometries, and as one can see in Figure \ref{fig:SPH_v_evol}, there was 
a rapid decrease in the number of fragments with initial $v>300\,\rm m\,s^{-1}$. 
The resulting histogram is again similar to that of the simple isotropic model. 

\subsection{Synthetic moons}

\begin{table}\vspace{-5mm}\centering
\caption{A comparison of the sizes and the orbital parameters (i.e. semimajor 
axis $a$, eccentricity $e$ and period $P$) of the observed moon of (624) Hektor 
as listed in Marchis et al. (2014), with the parameters of 
synthetic moons SPH I and SPH II captured in our SPH simulation of impact on 
the bilobed target.} 
\small
\renewcommand{\tabcolsep}{3pt}
\begin{tabular}{c|c|c|c|c|}	
\hline\
desig. & diam. [km] & $a$ [km] & $e$ & $P$ [days] \\
\hline\hline
observed & $12\pm3$ & $623.5\pm10$ & $0.31\pm0.03$ & $2.9651\pm0.0003$ \\
\hline
SPH I & 2.2 & 715 & 0.82 & 1.2 \\
SPH II & 2.7 & 370 & 0.64 & 0.4 \\
\hline
\vspace{2mm} 
\end{tabular}
\label{tab:moons}
\end{table}

In our simulation of the impact of basalt projectile on the bilobe-shape basalt 
target, we spotted two low-velocity fragments with original velocities 
$130\,\rm m\,s^{-1}$ and $125\,\rm m\,s^{-1}$, which were consequently 
captured as moons of the largest remnant. Their sizes and orbital parameters 
are listed in Table \ref{tab:moons}.

These satellites were captured on orbits with high eccentricities ($e=0.82$
and 0.64 respectively), which are much higher than the eccentricity of the 
observed moon determined by Marchis et al. (2014) ($e=0.31\pm0.03$). However, 
this could be partly caused by the fact, that we handed the output of 
(gravity free) SPH simulations to the gravitational N-body code after first 
100~s. Hence, fragments leaving the parent body could move freely without 
slowing down by gravity. More importantly, we do not account for any long-term 
dynamical evolution of the moons (e.g. by tides or binary YORP).

When compared to the observed satellite, the diameters of the synthetic moons 
are several times smaller. This is not too surprising, given that the results 
for satellite formation are at the small end of what can be estimated with our 
techniques (median smoothing length $h = 2.3\,{\rm km}$; satellite radius $r 
\simeq 1.2\,{\rm km}$). The size of captured fragments could also be dependent 
on impact conditions as different impact angles, impactor velocities and sizes 
(as is the case for scenarios of Moon formation) which we will analyze in 
detail in the future and study with more focused simulations.

\begin{figure*}
\centering
\begin{tabular}{cc}
\\
\includegraphics[width=9cm]{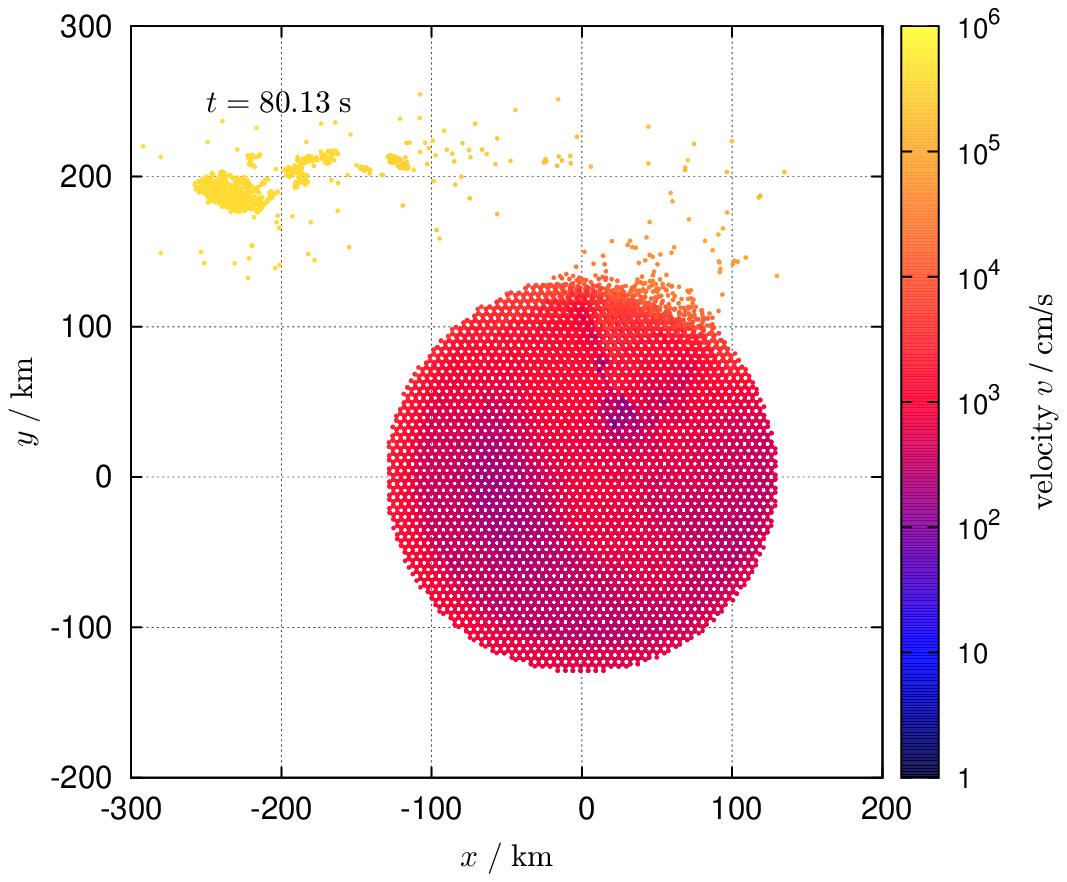} & 
\includegraphics[width=9cm]{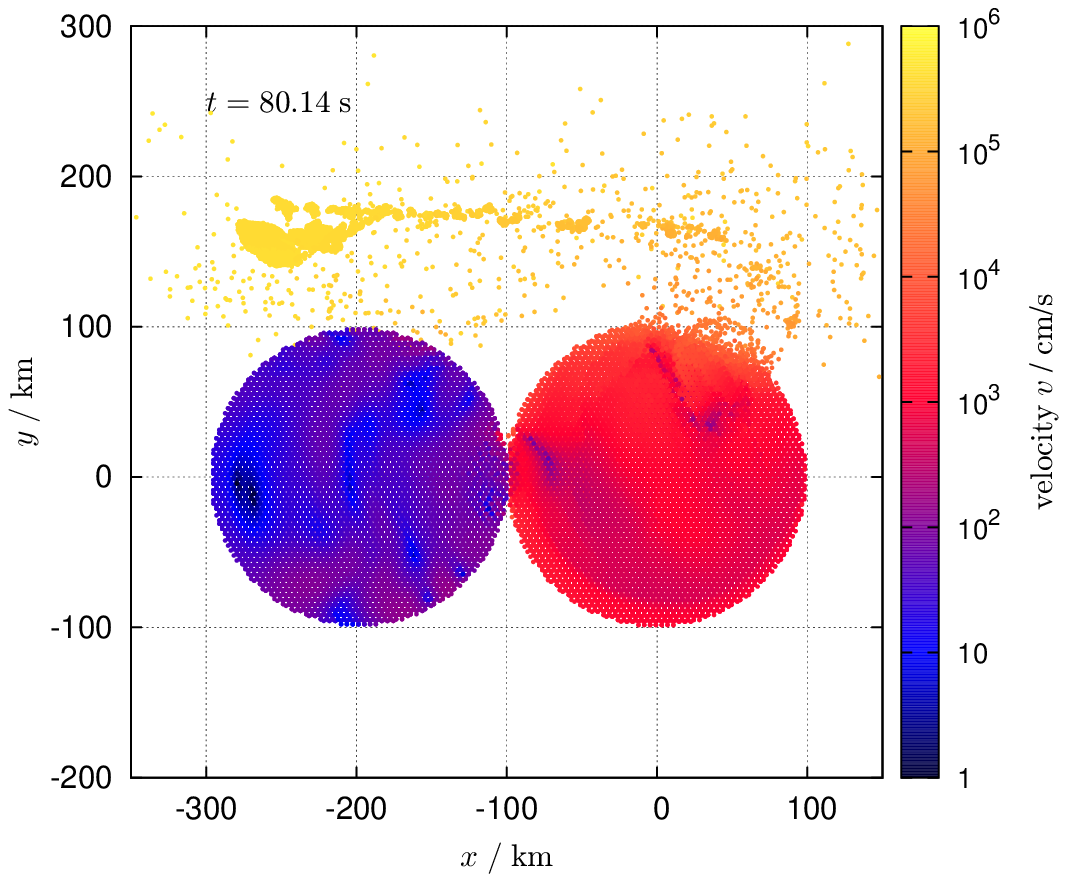} \\
\end{tabular}
\caption{A comparison of SPH simulations of a disruption of a single body 
(basalt) with diameter $D_{\rm target}=250\,\rm km$, by an impactor with the 
diameter $D_{\rm imp}=48\,\rm km$ (silicate ice) (\textit{left}) and 
a disruption of a bilobe basalt target, with $D_{\rm target}=198\,\rm km$   
for each sphere, by an impactor with $D_{\rm imp}=46\,\rm km$ (silicate ice) 
(\textit{right}). Time elapsed is $t=80.1\,\rm{s}$ in both cases. There are 
notable physical differences between the two simulations, especially in the 
propagation of the shock wave, which is reflected from free surfaces, the 
number of secondary impacts, or the fragmentation (damage) of the target. 
Nevertheless, the amount of ejected material and the resulting size-frequency 
distributions do not differ that much (cf. Figure \ref{fig:SPH_SFD}).}
\label{fig:SPH}
\end{figure*}

\section{Conclusions}\label{sec:conclusions}
In this paper, we updated the list of Trojans and their proper elements, 
what allowed us to update parameters of Trojan families and to discover a 
new one (namely $2001\,\rm UV_{209}$ in $L_5$ population). We focused on 
the Hektor family, which seems the most interesting due to the bilobed shape of 
the largest remnant with a small moon and also its D-type taxonomical 
classification, which is unique among the collisional families 
observed so far.

At the current stage of knowledge, it seems to us there are no major 
inconsistencies among the observed number of Trojan families and their 
dynamical and collisional evolution, at least in the current environment.

As usual, we ``desperately'' need new observational data, namely in the 
size range from 5 to 10 km, which would enable us to constrain the ages of 
asteroid families on the basis of collisional modeling and to decide between 
two proposed ages of Hektor family, 1~to~4~Gyr or 0.1~to~2.5~Gyr. 

As expected, there are qualitative differences in impacts on single and 
bilobed targets. In our setup, the shockwave does not propagate easily into 
the secondary, so that only one half the mass is totally damaged as one can see 
in Figure \ref{fig:SPH}. On the other hand, the resulting SFDs are not that 
different, as we would expect. 

Even so, there is a large parameter space, which is still not investigated 
(i.e. the impact geometry with respect to the secondary, 
secondary impacts, the position in the orbit). SPH simulations of impacts 
on bilobed or binary targets thus seem very worthy for future research.

Our work is also a strong motivation for research of 
disruptions of weak bodies (e.g. comets), better understanding the cometary 
disruption scaling law and also for experimental determination of material 
constants, which appear in the respective equation of state.

As a curiosity, we can also think of searching for the remaining projectile, 
which could be still present among Trojans on a trajectory substantially 
different from that of family. A substantial part of projectile momentum is 
preserved in our simulations, so we may 
turn the logic and we may assume the projectile most likely came from the 
Trojan region and then it should remain in this region too. 

\section*{Acknowledgements}
We thank Alessandro Morbidelli for his review which helped to improve the final 
version of the paper.

The work of MB was supported by the grant no. P209/13/01308S
and that of JR by P209/15/04816S of the Czech Science Foundation (GA CR). 
We acknowledge the usage of computers of the Stefanik Observatory, Prague.


\bsp

\label{lastpage}

\end{document}